\documentclass{aa}
\usepackage{graphicx}
\usepackage{amssymb}
\usepackage{amsmath}
\usepackage{natbib}
\usepackage{mathrsfs}
\usepackage{ulem}

\newcommand{\corot}{{\textsc{CoRoT}}}
\newcommand{\cible}{{HD~49385}}

\newcommand{\ind}[1]{_{\rm #1}}

\def\m2s2{\,m$^{2}$\,s$^{-2}$} %m2.s -2
\def\aov{\alpha\ind{ov}}

\newcommand{\expon}[1]{^{\rm #1}}
\newcommand{\eme}{\expon{th}}
\newcommand{\nuac}{\nu\ind{cross}}
\newcommand{\vaisala}{Brunt-V\"ais\"al\"a}
\newcommand{\cesam}{\textsc{cesam2k}}

\newcommand{\lsmean}{\langle\Delta\nu\rangle}

\newcommand{\gradmu}{\nabla\mu}

\newcommand\T{\rule{0pt}{2.6ex}}

\newcommand\B{\rule[-1.2ex]{0pt}{0pt}}

\begin{document}
\title{Constraints on the structure of the core of subgiants via mixed modes: the case of HD~49385}
%\subtitle{}
\titlerunning{Constraints on the structure of the core of HD~49385 via mixed modes}
\author{
S. Deheuvels\inst{1,2}
\and E. Michel \inst{1}
}

\institute{LESIA, UMR8109, Observatoire de Paris, Universit\'e Pierre et Marie Curie, Universit\'e Denis Diderot, CNRS, 5 Place Jules Janssen 92195 Meudon Cedex, France \\
              \email{sebastien.deheuvels@yale.edu}
              \and Department of Astronomy, Yale University, P.O. Box 208101, New Haven, CT 06520-8101, USA
}

%\offprints{S. Deheuvels\\ \email{sebastien.deheuvels@yale.edu}}

\date{Submitted ...}%; accepted March 16, 1997}

\abstract{The solar-like pulsator \cible\ was observed with the \corot\thanks{Based on data obtained from the CoRoT (Convection, Rotation and planetary Transits) space mission, developed by the French Space agency CNES in collaboration 
with the Science Programs of ESA, Austria, Belgium, Brazil, Germany and Spain.} satellite during 137 days. The analysis of its oscillation spectrum yielded precise estimates of the mode frequencies over nine radial orders and pointed out some unusual characteristics: there exist some modes outside the identified ridges in the \'echelle diagram and the curvature of the $\ell=1$ ridge significantly differs from that of the $\ell=0$ ridge.}
{We first search for stellar models reproducing the peculiar features of the oscillation spectrum of \cible. Having shown that they can be accounted for only by a low-frequency $\ell=1$ avoided crossing, we investigate the information which is brought by the mixed modes about the structure of the core of \cible.}
{We propose a toy-model to study the case of avoided crossings with a strong coupling between the p-mode and g-mode cavities in order to establish the presence of mixed modes in the spectrum of \cible. We then show that traditional optimization techniques are ill-suited for stars with mixed modes in avoided crossing. We propose a new approach to the computation of grids of models which we apply to \cible.}
{The detection of mixed modes leads us to establish the post-main-sequence status of \cible. The mixed mode frequencies suggest a strong coupling between the p-mode and g-mode cavities. As a result, we show that the amount of core overshooting in \cible\ is either very low ($0<\aov<0.05$) or moderate ($0.18<\aov<0.20$). The mixing length parameter is found to be significantly lower than the solar one ($\alpha\ind{CGM}=0.55\pm0.04$ compared to the solar value $\alpha_{\odot}=0.64$). Finally, we show that the revised solar abundances of Asplund give a better agreement than the classical ones of Grevesse \& Noels. At each step, we investigate the origin and meaning of these seismic diagnostics in terms of the physical structure of the star.}
%We thus establish that the coupling between the cavities strongly depends on the stellar mass.}
%reasons why the indication of a strong coupling between the cavities leads to such constraints on the stellar structure.}
%
{The subgiant \cible\ is the first star for which a thorough modeling was led, trying to reproduce all the properties of an avoided crossing. It gave the opportunity to show that the study of the coupling between the cavities in these stars can provide valuable contribution to the questions of core overshooting, the efficiency of convection and the abundances of heavy elements in stars.}

%\abstract{bla}
%%
%{bla}
%%
%{bla}
%%
%{bla}

\keywords{Stars: oscillations -- Stars: interiors -- Stars: evolution -- Stars: individual: HD~49385}

%we are able to constrain the mass of the star with a 3.5 \% precision.
%We find that only certain types of models reproduce the strong coupling between the p-mode and g-mode cavities which is suggested by the mixed mode frequencies. As a result 
%The mixed mode frequencies suggest a strong coupling between the p-mode and g-mode cavities. We show that this brings strong constraints on the amount of core overshooting in \cible, which can be either very low ($0<\alpha\ind{conv}<0.05$) or moderate ($0.18<\alpha\ind{conv}<0.20$). We also obtain 
%

\maketitle
%________________________________________________________________
\section{Introduction \label{sect_intro}}

%\begin{itemize}
%\item Evolved stars
%\item \corot\
%\item Originality of this study
%\end{itemize}

%As a star evolves, the increasing density in the inner regions generates an increase of the g-mode frequencies. At the same time, the p-mode frequencies decrease due to the expansion of the star. The frequency range of the g modes and that of the p modes eventually overlap, making it possible for a g mode and a p mode of like degree to have frequencies very close to each other. When this happens, instead of actually meeting, the two modes undergo an 'avoided crossing' resulting in them exchanging natures. During this phenomenon, both modes present a mixed character: they behave as g modes in the inner regions and as p modes in the envelope. 

\begin{figure}
\includegraphics[width=8.5cm]{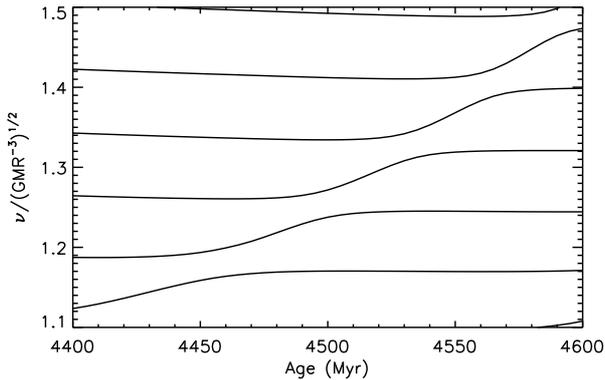}
\caption{Evolution of the eigenfrequencies of $\ell=1$ modes as a function of the age for a $1.3\,M_{\odot}$ model. The frequencies are normalized by the square root of the mean density $\sqrt{GM/R^3}$, where $M$ and $R$ are the stellar mass and radius. 
\label{fig_evol_mm_intro}}
\end{figure}

The existence of mixed modes in the spectrum of stars has been first suggested by \cite{scuflaire74}. By studying non-radial oscillations of highly condensed polytropes, he found waves behaving both as gravity waves in the center, and as acoustic waves in the envelope. Such modes were later discovered in the spectrum of $10\,M_{\odot}$ models by \cite{osaki75}, who established that they are associated with \textit{avoided crossings} between g modes and non-radial p modes. %Focusing on the evolution of the eigenfrequencies of $\ell=2$ modes as a function of the age in his models, 
He showed that whenever the frequencies of two of modes of like degree got close to each other, the modes would avoid each other and exchange natures instead of actually crossing (see Fig. \ref{fig_evol_mm_intro}). During this exchange, they have a mixed character, like the ones found by \cite{scuflaire74}. This phenomenon is caused by the evanescent zone which separates the p-mode cavity from the g-mode cavity and introduces a coupling between them. \cite{aizenman77} gave evidence of that by decoupling the two cavities and showing that the modes do cross in this case.

Several studies have stressed the great expected potential of avoided crossings in terms of asteroseismic diagnostics. They indeed provide an estimate of the frequency of the g mode they involve. This is crucial since the g-mode frequencies are determined by the profile of the \vaisala\ frequency in the core and for stars massive enough to have a convective core, this quantity depends to a great extent on the structure of the chemically inhomogeneous zone generated by the withdrawal of the core. This led to the natural idea that the frequency of mixed modes could be used as a means of constraining the amount of overshooting at the boundary of convective cores (e.g. \citealt{dziembowski91}).

Until now, there have been very few detections of stellar oscillation modes in avoided crossings. Mixed modes were observed in the subgiant star $\eta$ Boo (\citealt{kjeldsen95b}). They led to favor a post-main-sequence status for the star (\citealt{dimauro04}, \citealt{carrier05}) and to set an upper limit for the overshooting, but the data were not precise enough to further constrain the internal structure of the star. The recent development of space missions \corot\ (\citealt{baglin06}) and Kepler (\citealt{koch10}), by providing long quasi uninterrupted time series of high precision photometric data, has opened new opportunities for the detection of mixed modes.

In this paper, we investigate the case of the G0-type star \cible, which was observed with the satellite \corot\ over a period of 137 days between October 2007 and March 2008. The analysis of the time series has shown that the star exhibits solar-like oscillations (\citealt{analyse_49385}, further referred to as D10) and the authors could unambiguously identify modes of degree $\ell=0,1,2$ over nine radial orders. Precise estimates of the mode properties (frequencies, linewidths, amplitudes) were obtained by fitting Lorentzian profiles on the observed spectrum. The authors also pointed out several striking features of the oscillation spectrum of \cible. They detected significant peaks which do not follow the expected pattern of high-radial-order p modes. Based on the low value of the observed surface gravity ($\log g = 4.00\pm0.06$, D10) which indicates that the star is certainly an evolved object, they mentioned the possibility that some of these peaks might be the signature of mixed modes. The analysis of the oscillation spectrum also showed that the curvature of the $\ell=1$ ridge in the \'echelle diagram unexpectedly differs from that of the $\ell=0$ ridge at low frequency.

We here perform a modeling of \cible\ based on the spectroscopic and seismic constraints derived for the star by D10. We present in Sect. \ref{sect_prelim_model} a preliminary modeling of the star which shows that main-sequence models fail to reproduce the peculiar curvature of the observed $\ell=1$ ridge. We then investigate the possibility that this feature might be caused by the presence of mixed modes in the oscillation spectrum of \cible. Until now, theoretical studies led on avoided crossings all made the assumption that only two modes are involved and they neglected the contributions of the other modes (e.g. \citealt{christensen81}, \citealt{gabriel80}). We study in Sect. \ref{sect_analogy} the case of avoided crossings when the coupling between the p-mode cavity and the g-mode cavity is too strong to consider this phenomenon as a two-mode only interaction and we show that they can generate a distortion of the ridge comparable to the observed one. This leads us to show that the seismic properties of \cible\ can only be accounted for by the existence of an $\ell=1$ avoided crossing. We then point out in Sect. \ref{sect_nuac} the limitations of traditional modeling techniques when studying stars with avoided crossings and we propose a method to remedy this. This method is applied to search for optimal models of \cible\ in Sect. \ref{sect_optim}, with the aim to determine the constraints that the frequencies of the mixed modes bring on the structure of the inner regions of \cible. In Sect. \ref{sect_discussion}, we discuss these results and investigate their meaning in terms of internal structure and physical processes.

\section{Characteristics of \cible\ and first step modeling \label{sect_prelim_model}}

%We now apply the results obtained in Sect. \ref{sect_analogy} to investigate the possible existence of mixed modes in the spectrum of the solar-like pulsator \cible. 

%The G0-type star \cible\ was one of the targets of the second 'Long Run' in the Seismo field of the space mission \corot. It was observed during 137 days with a duty cycle of 88.2\%. To complete 

%is a G0-type star which was recently observed in photometry by the space telescope \corot\ over a 'Long Run' (137 days) with a duty cycle of 88.2\%. 

%which benefited from both spectroscopic ground observations with 

\subsection{Observational constraints}

We first give a brief overview of the observational constraints which were derived for this star in previous studies. %before presenting a preliminary modeling of the object. 

\subsubsection{Surface constraints \label{sect_surf_constr}}

\begin{table}
  \begin{center}
  \caption{Fundamental parameters of \cible\ measured from spectroscopic and photometric observations by D10 (\textit{top}) and estimated using the observed value of $\Delta\nu$ and $\nu\ind{max}$ (\textit{bottom}). \label{tab_fund}}
\begin{tabular}{ l  c }
\hline \hline
\multicolumn{2}{l}{Measured Parameters} \\
\hline
\T $T\ind{eff}$ (K) & $ 6095 \pm65 $  \\
$(\log g)\ind{spectro} $   & $ 4.00 \pm 0.06$ \\
$[Z/X]$ (dex)$^{(\star)}$ & $+0.09 \pm 0.05$ \\
\B $\log(L/L_{\odot})$ & $0.67\pm0.05$ \\
\hline \hline
\multicolumn{2}{l}{Parameters estimated from $\Delta\nu$ and $\nu\ind{max}$} \\
\hline
\T $M/M_{\odot}$ & $1.31\pm0.12$ \\
$R/R_{\odot}$ & $1.96\pm0.07$ \\
\B $(\log g)\ind{seismo}$ & $3.97\pm0.02$ \\
\hline
\end{tabular}
\end{center}
{\small \textbf{Notes.} $^{(\star)}$ The metallicity is  defined as $[Z/X]\equiv\log\left[(Z/X)/(Z/X)_{\odot}\right]$.}
\end{table}

The surface observables of \cible\ were derived by D10 based on a detailed analysis of two high-quality spectra obtained with the NARVAL spectrograph and on the Hipparcos measurement of the star's parallax. We here use their results which are recalled in Table \ref{tab_fund}.

%based on the analysis of two high-quality spectra obtained with the NARVAL spectrograph mounted on the 2-m Bernard Lyot Telescope at the Pic du Midi Observtory. Using the semi-autmoatic software VWA (\citealt{bruntt09}), the authors obtained estimates of the stellar temperature, surface gravity and metallicity, which can be found in Table \ref{tab_fund}. 

%Using the Hipparcos parallax of the star ($\pi=13.91\,\pm\,0.76$ mas, \citealt{vanleeuwen07}) along with the estimate of its apparent magnitude ($m\ind{V}=7.39$, \citealt{hauck98}) and a bolometric correction derived from \cite{bessell98}, the authors obtained a stellar luminosity of $\log(L/L_{\odot})=0.67\,\pm\,0.05$ for \cible. 

%The apparent magnitude of \cible\ is $m\ind{V}=7.39$ (\textit{uvby} catalog, \citealt{hauck98}). Using the Hipparcos parallax of the star, $\pi=13.91\,\pm\,0.76$ mas (\citealt{vanleeuwen07}), we obtain its absolute visual magnitude, $M\ind{V}=3.11\pm0.12$. With a bolometric correction of $BC\ind{V}=-0.029\,\pm\,0.006$, obtained by interpolating in the grid provided by \cite{bessell98}, we derive a luminosity of $\log(L/L_{\odot})=0.67\,\pm\,0.05$ for \cible. 

\subsubsection{Seismic constraints \label{sect_sism_constr}}

The star \cible\ is a solar-like pulsator which has been observed with the \corot\ satellite during 137 days.
%pulsator \cible\ is a G0-type star which was one of the targets of the second Long Run of the space mission \corot. It was observed during 137 days with a duty cycle of 88.2\%. 
%Solar-like oscillations were detected in the power spectrum of the lightcurve, in the range [$700, 1300$] $\mu$Hz (D10). 
The ridges of degrees $0\leqslant\ell\leqslant2$ were unambiguously identified in the \'echelle diagram of the power spectrum (D10). The frequencies of these modes were determined over nine radial orders by fitting Lorentzian profiles on the detected modes, using a maximum likelihood estimation. 
%They are added to the classical constraints given in Sect. \ref{sect_surf_constr} to form the set of observables which will be used to constrain our models. 
%We note that D10 mentioned the possible detection of $\ell=3$ modes. They also found the frequency of 

The star has a mean large separation of $\lsmean\ind{obs}=56.3\pm0.5\,\mu$Hz and the frequency at the maximum of the signal is $\nu\ind{max}=1010\pm10\,\mu$Hz (the error bars on these measures were determined based on the method prescribed by \citealt{mosser_autocor}). These values can be used to derive a first estimate of the stellar mass and radius using scaling relations. \cite{gai11} recently noted that since the errors in these estimates of the mass and radius are positively correlated, the error we obtain on the quantity $M/R^2$ and thus on $\log g$ is small. We reported in Table \ref{tab_fund} the estimated values of $M$, $R$ and $\log g$ which will later be confronted to the ones of our best fitting models. We notice that the value we obtain for $\log g$ is consistent with the spectroscopic measure of this quantity and has a much lower error, as expected.

\begin{figure}
\includegraphics[width=8.5cm]{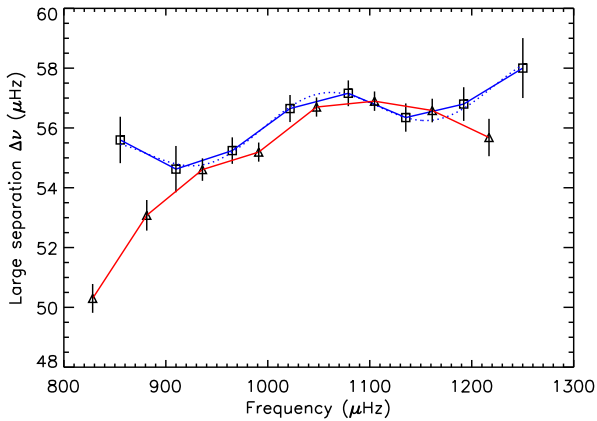}
\caption{Profile of the $\ell=0$ (squares and blue line) and $\ell=1$ (triangles and red line) observed large separation represented with 1-$\sigma$ error bars. The blue dotted line shows the result of the least-square-fit to a sinusoid of the $\ell=0$ large separation (see text).
\label{fig_large_sep}}
\end{figure}

A clear oscillation can be detected in the profile of the $\ell=0$ large separation, as can be seen in Fig. \ref{fig_large_sep}. Following the procedure proposed by \cite{roxburgh09b}, we estimated a period of $230\pm30\,\mu$Hz for this oscillation. This indicates an acoustic depth of the glitch responsible for the oscillation of either $\tau\ind{glitch}/\tau_{\star}=0.76\pm0.04$ or $\tau\ind{glitch}/\tau_{\star}=0.24\pm0.04$, where $\tau_{\star}$ is the total acoustic radius of the star (we recall that two glitches at acoustic depths of $\tau$ and $\tau_{\star}-\tau$ result in an oscillation of the eigenfrequencies with the same period). 
%A more precise estimate could be provided following the method proposed by \cite{mazumdar10}, 
%could apply the method proposed by \cite{mazumdar10} to try to determine in a model-independant way the acoustic depth of the glitch responsible for this oscillation. However, such a study is out of the scope of the present paper, and we not that the small number of available points would certainly limit the conclusions which would be drawn from it. For future comparisons, we followed the procedure advocated by \cite{roxburgh09b} to remove the long-period trend in the profile of the $\ell=0$ large separation and we then performed a least-square fit to estimate the period of the oscillation in the residual. We found a period of $230\pm30\,\mu$Hz, indicating an acoustic depth of about $\tau\ind{glitch}/\tau_{\star}=0.76\pm0.04$ or $\tau\ind{glitch}/\tau_{\star}=0.24\pm0.04$, where $\tau_{\star}$ is the total acoustic radius of the star (we recall that two glitches at acoustic depths of $\tau$ and $\tau_{\star}-\tau$ respectively result in an oscillation of the eigenfrequencies with the same period).

The analysis of the oscillation spectrum of \cible\ led D10 to remark several unaccounted for features. First, contrary to what is expected for main sequence solar-like pulsators, the profile of the $\ell=1$ large separation significantly differs from that of the $\ell=0$ large separation, especially in the low-frequency part of the observations (see Fig. \ref{fig_large_sep}). In an \'echelle diagram, this translates into a growing difference between the curvatures of the $\ell=0$ and $\ell=1$ ridges at low frequency. Secondly, several peaks were found to be significant even though they lie outside of the identified ridges. One of them, the peak labeled as $\pi_1$ by the authors, has a posterior probability of being due to noise as low as $10^{-5}$. It lies close to the $\ell=0$ ridge but not in its direct continuity and it would cause an abrupt step in the $\ell=0$ large separation profile if it were identified as a radial mode. One of the challenges of the modeling of \cible\ will be to understand these peculiarities. We address this question in the following section.

\subsection{Preliminary modeling of HD~49385}

\subsubsection{Properties of the models}

All the models are computed with the stellar evolution code \cesam\ (\citealt{cesam}) and the mode frequencies are derived from them using the Li\`ege Oscillation Code (LOSC, \citealt{losc}). We used the OPAL equation of state and opacity tables as described in \cite{lebreton08}. The nuclear reaction rates are computed using the NACRE compilation (\citealt{angulo99}). The atmosphere is described by Eddington's grey law and is connected to the envelope at an optical depth of $\tau=10$ to ensure the validity of the diffusion approximation (\citealt{morel94}). We computed models using alternately the mixture of heavy elements of \cite{grevesse93}, further referred to as GN93, and the revised mixture of \cite{asplund05}, referred to as AGS05.

The convective regions are treated using the Canuto-Goldman-Mazzitelli (CGM) formalism (\citealt{canuto96}). It involves a free parameter, the mixing length, described as a fraction $\alpha\ind{CGM}$ of the pressure scale height. A calibration on the Sun gives $\alpha\ind{CGM}=0.64$ (\citealt{samadi06}). In this work, the mixing length will be considered as a free parameter. 

The radius $R\ind{cc}$ of the convective core is determined by the Schwarzschild criterion. Overshooting can be included in the models as an extension of the motion of convective eddies over a distance $d\ind{ov}$ outside the core, which is defined as
\begin{equation}
d\ind{ov} \equiv \alpha\ind{ov} \min\left( R\ind{cc}, H_P \right)
\end{equation}
where $H_P$ is the pressure scale height. In this study, we consider an instantaneous mixing in the overshooting region, meaning we assume that the time scale of the mixing is much smaller than the time scale of the evolution of the star. In this case the overshooting zone is fully mixed. The temperature gradient is taken to be equal to the adiabatic gradient.

Our models are computed neglecting microscopic diffusion, in order to limit the computational time of our grids of models. However, the effect of microscopic diffusion on our conclusions is studied in Sect. \ref{sect_discussion}.

\subsubsection{Comparison criterion between models and observations \label{sect_comp_mod_obs}}

As is usually done, we compare the stellar models and the observations by computing the merit function $\chi^2$ defined as
\begin{equation}
\chi^2\equiv \sum_{i=1}^N \left[ \frac{\mathcal{O}_i\expon{mod}-\mathcal{O}_i\expon{obs}}{\sigma_i\expon{obs}} \right]^2 \label{eq_chi2}
\end{equation}
where $\mathcal{O}_i\expon{obs}$, $i=1,N$ represent the $N$ observables selected to constrain the models, $\sigma_i\expon{obs}$ their error bars and $\mathcal{O}_i\expon{mod}$ the values of these parameters for the computed models.

To constrain the models, we used the classical observables $T\ind{eff}$ and $L/L_{\odot}$. We note that the measured value of $\log g$ could also be used. However its value is not well enough constrained by spectroscopic measurements to allow it to discriminate among the different models and we will therefore no longer mention it in the following. The set of observables is completed by the estimates of the p-mode frequencies of degrees $0\leqslant\ell\leqslant2$ obtained by D10. We note that the authors also mentioned the possible detection of two $\ell=3$ modes (modes $\pi_2$ and $\pi_3$) and proposed estimates of their frequencies. However, given their weak signal-to-noise ratio, we chose not to include them in our set of observables. 
%\textbf{Besides, based on the best models which will be obtained for \cible\ in Sect. \ref{sect_optim}, the modes $\pi_2$ and $\pi_3$ are about 2 $\mu$Hz too close to the $\ell=1$ ridge to correspond to $\ell=3$ modes, which sheds doubt on their identification.} 
Additionally, one of the $\ell=2$ modes was found to overlap the closest radial mode ($\nu_{0,14}=855.3\;\mu$Hz). Since none of all the models we computed in this study present this particularity, we supposed that this phenomenon is due to the lower signal-to-noise ratio of $\ell=2$ modes around the edges of the frequency domain of the oscillations. This $\ell=2$ mode was therefore not included among the observables $\mathcal{O}_i\expon{obs}$.

To compare the eigenfrequencies between models and observations, we know that it is necessary to correct them from the effects of our improper modeling of the structure in the surface layers (e.g. \citealt{christensen97}). \cite{kjeldsen08} found that in the case of the Sun, the differences between the observed frequencies and those of the best solar model were well approximated by a power law of the form
\begin{equation}
\nu\expon{obs}_{n,\ell}-\nu\expon{best}_{n,\ell} = a\left(\frac{\nu\expon{obs}_{n,\ell}}{\nu_0}\right)^b \label{eq_nse}
\end{equation}
where $\nu_0$ is the frequency of the maximum of the signal. The authors tested this law on a few other solar-like pulsators and found a reasonable agreement. For this purpose, they assumed that the exponent found in the solar case ($b=4.90$) could also be used for other targets. We applied a correction of this type on the frequencies of our models. Only our case is somewhat different from that of \cite{kjeldsen08}: our models are computed using the \cesam\ code and we treat convection with the formalism of \cite{canuto96} whereas \cite{kjeldsen08} use ASTEC and describe convection using the traditional mixing length theory. It would therefore be irrelevant to use the same exponent as the one found by the authors in the solar case. We computed a solar model matching the set of solar eigenfrequencies found by \cite{gelly02}. We then fitted the power law given by Eq. \ref{eq_nse} on the differences between the observed frequencies and those of our solar model. We obtained a somewhat smaller exponent ($b=4.25$) which is used to correct the frequencies in the following. We also note that the modes which have a mixed behavior are expected to be less sensitive to the surface effects. To take this into account, for non-radial modes, the surface correction given by Eq. \ref{eq_nse} was multiplied by a factor $Q_{n,\ell}^{-1}$, where $Q_{n,\ell}$ corresponds to the ratio between the inertia of the mode and the inertia of the closest radial mode, as prescribed by \cite{asteroseismology}.

%, we use the formalism of \cite{canuto96} to describe convection

%We also used the seismic constraints derived by D10. To compare the eigenfrequencies between models and observations, 

%we applied a correction to compensate the improper modeling of the surface layers. Based on the solar data, 

%first applied a correction of surface effects following the procedure proposed by \cite{kjeldsen08}. We made the assumption that 

%The models are constrained using the classical observables $T\ind{eff}$ and $L/L_{\odot}$. We note that the measured value of $\log g$ can be used also 

%To compare the models with the observations, we use both the classical constraints available for this star ($T\ind{eff}$, $L/L_{\odot}$) and the seismological constraints (frequencies of modes of degrees $0\leqslant\ell\leqslant 2$)

\subsubsection{Evolutionary status of HD~49385}

\begin{figure}
\includegraphics[width=8.5cm]{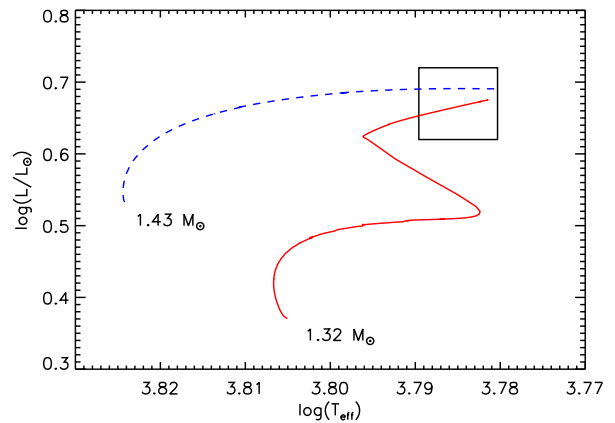}
\caption{Evolutionary tracks of two models fitting the position of \cible\ in the HR diagram. The observed values of the effective temperature and luminosity of \cible\ are indicated within 1-$\sigma$ error bars by the box. The blue dashed line corresponds to a MS model, and the red solid line to a PoMS model.
\label{fig_hr}}
\end{figure}

The low value of the spectroscopic $\log g$ obtained for \cible\ suggests that the star is evolved. We can find models both in the main sequence (hereafter MS) and post main sequence (PoMS) stage, which fit the position of \cible\ in the Hertzsprung-Russell (HR) diagram (see Fig. \ref{fig_hr}). We are facing a problem which is quite commonplace for evolved objects: a degeneracy of MS models and PoMS models which results in an uncertainty on the evolutionary stage of the studied object (see \textit{e.g.} Procyon A in \citealt{barban99}, \citealt{provost06}, $\eta$ Boo in \citealt{dimauro03}).

The internal structure of these two families of models is very different. The main sequence models are close to the exhaustion of their hydrogen reserves in the center ($X\ind{c}\sim0.1$) and they have a small convective core spreading over about $5\%$ of the stellar radius. The post main sequence models are burning hydrogen in a thin layer located above the limit of the convective core which existed during the main sequence stage. There are no longer any convective regions in the center of these models.

\subsubsection{Main sequence models}

\begin{figure}
\includegraphics[width=8.5cm]{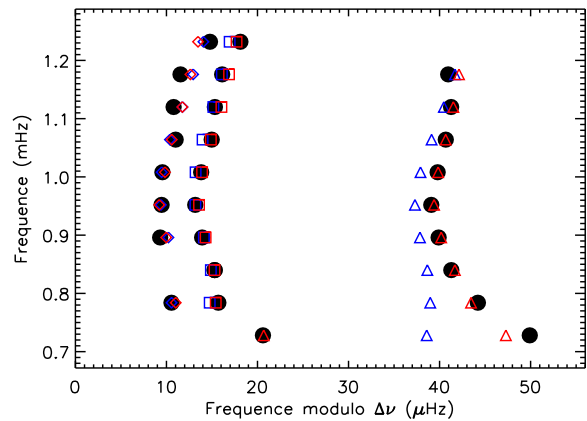}
\caption{\'Echelle diagram of the power spectrum of \cible\ derived from 137 days of \corot\ data, folded with a mean large separation $\Delta\nu=56.3\,\mu$Hz. The frequencies of a MS model and a PoMS model are overplotted in blue and red, respectively. The models have been computed to reproduce the observed $\ell=0$ large separation as well as the position of the star in the HR diagram and the PoMS has an $\ell=1$ avoided crossing at low frequency. Squares represent $\ell=0$ modes, triangles $\ell=1$ modes and diamonds $\ell=2$ modes.
\label{fig_ech_superpose}}
\end{figure}

We computed a grid of main sequence models with varying masses, ages, helium abundances, mixing length parameters and metallicities. We found models which fit quite well the observed variations of the $\ell=0$ large separation and the position of the star in the HR diagram. The \'echelle diagram of one of them is represented in Fig. \ref{fig_ech_superpose}. However, none of the computed MS models are able to reproduce the peculiar curvature of the $\ell=1$ ridge which we mentioned before. 
%As a result, there is a major disagreement between MS models and observations for the low frequency profiles of the $\ell=1$ large separation and the seismic index $\zeroun$. The small separation $\zerodeux$ is also not well fitted by MS models which produce values of this seismic index systematically smaller than the observed ones. 
The values of the $\chi^2$ function we obtain for MS models are all above 2000. This yields a reduced $\chi^2$ above 100, which indicates a very poor match with the observations. The high value of $\chi^2$ for MS models is almost entirely due to the contribution of the $\ell=1$ modes at low frequency (the four $\ell=1$ modes with the lowest frequencies account for more than 90\% of the $\chi^2$ value for the best MS models). 

Besides, no MS model can explain the presence of the peak $\pi_1$. It cannot correspond to the signature of a mixed mode, since MS models which reproduce the surface observables of \cible\ are not evolved enough to have mixed modes in the frequency domain of the observations. This peak could not either be identified as an $\ell=0$ mode since the closest radial mode in the models lies about 5 $\mu$Hz away from it (i.e. at more than $20\sigma$). 
%There exists quite a number of disagreements between the seismic parameters of MS models and those which were observed for \cible.

%The seismic parameters  and the evolutionary track of the best MS model we found are shown in Fig. \ref{fig_ech_superpose}. We can see on this figure that the observed variations of the $\ell=0$ large separation are quite well fitted.

\subsubsection{Post main sequence models}

Contrary to MS models, some PoMS models which fit the surface parameters of \cible\ have $\ell=1$ mixed modes in the frequency domain of the observations. We have considered the possibility that the difference we observe between the curvatures of the $\ell=1$ ridge and the $\ell=0$ ridge might be caused by an $\ell=1$ avoided crossing. So far, the studies which were led about avoided crossings made the assumption that only two modes were involved and they neglected the contributions of the other modes (e.g. \citealt{christensen81}, \citealt{gabriel80}). In the next section, we investigate the case when the coupling between the cavities is such that this hypothesis is no longer valid and we try to determine the effect it has on the curvature of the ridge.

\section{Avoided crossings with strong coupling \label{sect_analogy}}

\subsection{Analogy with harmonic oscillators}

%We start by presenting a simple analogy which illustrates the main aspects of mixed modes in avoided crossing and will provide helpful insights for the subsequent modeling of \cible. This analogy has been presented in \cite{rome} but we recall it here for the sake of clarity.

%The fact that an avoided crossing occurs between two modes of same degree $\ell$ and whose eigenfrequencies are close to each other is due to the coupling that exists between the p-mode cavity and the g-mode cavity inside the star. Up until now, the studies which were led about this phenomenon made the assumption that  only two modes were involved and neglected the contributions of the other modes (e.g. \citealt{christensen81}, \citealt{gabriel80}). We investigated the cases where this hypothesis is no longer valid by 

%We here propose an extension of a simple analogy proposed by \cite{lecturenotesJCD} based on the work of \cite{vonneuman29}. This analogy illustrates the main aspects of mixed modes in avoided crossing 

We here develop an analogy introduced by \cite{rome} to gain insights on the characteristics of avoided crossings involving $n$ coupled modes. We recall its main points, for the sake of clarity, and we add some details about the choice of the coupling term.

%\cite{lecturenotesJCD}, based on the work of \cite{vonneuman29}, proposed a simple analogy which captures the main aspects of avoided crossings between two modes. We here present an extension of this analogy to the case of $n$ coupled modes, which will provide helpful insights for the subsequent modeling of \cible. 

\subsubsection{Avoided crossing with two modes}

\cite{lecturenotesJCD}, based on the work of \cite{vonneuman29}, proposed a simple analogy which captures the main aspects of avoided crossings between two modes. He considers the two cavities of the star as two coupled harmonic oscillators $y_1(t)$ and $y_2(t)$, responding to the following system of equations:
\begin{eqnarray}
\frac{\hbox{d}^2y_1(t)}{\hbox{d}t^2} & = & -\omega_1(\lambda)^2y_1+\alpha y_2 \label{eqdiff_1} \\ 
\frac{\hbox{d}^2y_2(t)}{\hbox{d}t^2} & = & -\omega_2(\lambda)^2y_2+\alpha y_1 \nonumber
\end{eqnarray}
where $\alpha$ is the coupling term between the two oscillators and $\omega_1(\lambda)$, $\omega_2(\lambda)$ are the eigenfrequencies of the uncoupled oscillators (particular case where $\alpha=0$). They are chosen to depend on a parameter $\lambda$, used to model the change of the dimensions of the cavities as the star evolves. We suppose that for a certain $\lambda=\lambda_0$, the frequencies of the uncoupled oscillators cross, i.e. $\omega_1(\lambda_0)=\omega_2(\lambda_0)\equiv\omega_0$.

\begin{figure}
\includegraphics[width=8.5cm]{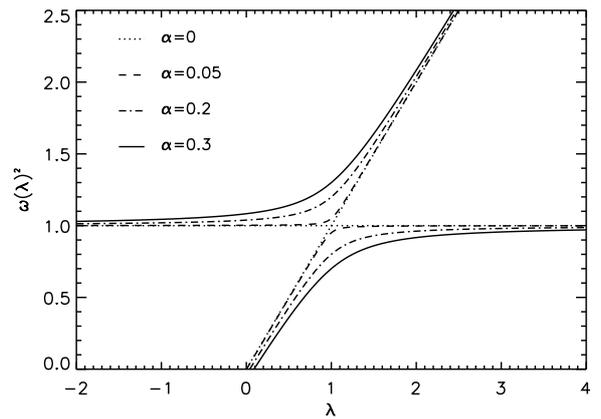}
\caption{Variations of the eigenfrequencies $\omega_{\pm}$ of the system with parameter $\lambda$, for different values of the coupling term $\alpha$.
\label{fig_cross12}}
\end{figure}

%The solutions of Eq. \ref{eqdiff_1} and \ref{eqdiff_2} can be written under the form $y_j(t)=c_j \exp(-i\omega t)$ ($j=1,2$), where the coefficients $c_j$ need to be determined. By inserting this into Eq. \ref{eqdiff_1} and \ref{eqdiff_2}, we find that the eigenfrequencies of the system are obtained by solving the eigenvalue problem $AC=\omega^2 C$, where
%\begin{equation}
%A=
%\begin{bmatrix}
%\omega_1^2 & -\alpha \\
%-\alpha & \omega_2^2
%\end{bmatrix}
%\;\;\;\;\; \hbox{and} \;\;\;\;\; C=
%\begin{bmatrix}
%c_1 \\
%c_2
%\end{bmatrix}.
%\end{equation}

Solving Eq. \ref{eqdiff_1}, we obtain the two following solutions:
\begin{equation}
\omega_{\pm}^2=\frac{\omega_1^2+\omega_2^2}{2}\pm\frac{1}{2} \sqrt{ (\omega_1^2-\omega_2^2)^2+4\alpha^2 } \label{sol12}
\end{equation}
If the coupling term $\alpha$ is very small compared to the difference between the eigenfrequencies ($\alpha\ll |\omega_1^2-\omega_2^2|$), then the eigenfrequencies of the system are close to $\omega_1$ and $\omega_2$. If, on the contrary, $|\omega_1^2-\omega_2^2|\ll \alpha_{1,2}$, then the eigenfrequencies can be approximated by 
\begin{equation}
\omega_{\pm}^2=\omega_0^2\pm \alpha. \label{eq_near_AC}
\end{equation}
%It is clear that the two oscillators "avoid" the frequency $\omega_0$, and the larger $\alpha$, the further away of $\omega_0$ they are during the avoided crossing.
To analyze these solutions, we choose the eigenfrequencies of the uncoupled oscillators $\omega_1(\lambda)$ and $\omega_2(\lambda)$ such that they simulate a p mode and a g mode. By reference to Fig. \ref{fig_evol_mm_intro}, we impose $\omega_1(\lambda) = 1$ and $\omega_2(\lambda) = \lambda$. In that case, an avoided crossing occurs in the system around $\lambda_0=1$. The variations of the two eigenfrequencies $\omega_{\pm}(\lambda)$ are shown in Fig. \ref{fig_cross12}. The two modes clearly exchange natures during the avoided crossing and the intensity of the phenomenon depends on the strength of the coupling between the oscillators.

\subsubsection{Avoided crossing with $n$ modes}

\begin{figure}
\includegraphics[width=8.5cm]{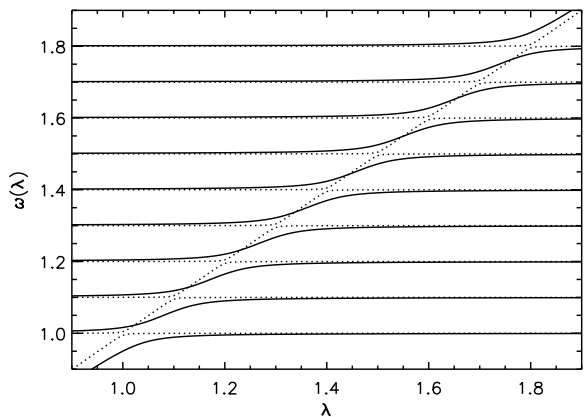}
\caption{Variations of the eigenfrequencies of $(n-1)$ $p$ modes coupled to a $g$ mode which undergoes avoided crossings with the $p$ modes (here, $n=10$).
The dashed lines correspond to a "weak coupling" ($\gamma=0.01$), and the full lines to a "strong coupling" ($\gamma=0.06$).
\label{fig_cross_n}}
\end{figure}

\begin{figure}
\begin{center}
\includegraphics[width=8.5cm]{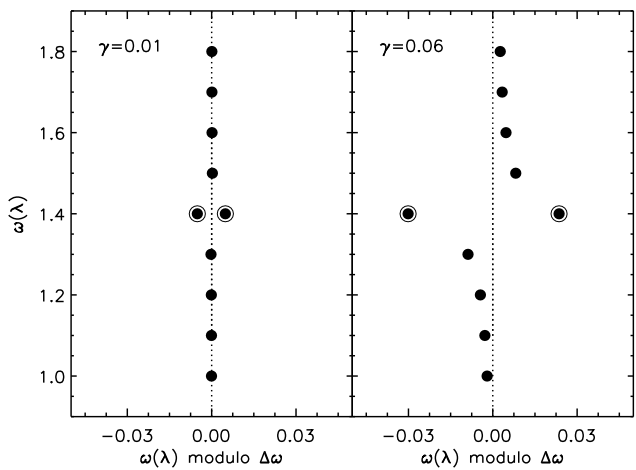}
\end{center}
\caption{\'Echelle diagrams of the eigenfrequencies of 10 coupled harmonic oscillators, 
%at three different "times" $\lambda_1<\lambda_2<\lambda_3$ 
at a given "time" $\lambda$ in the vicinity of an avoided crossing. The left panel presents the case of a weak coupling ($\gamma=0.01$) and the right panel the case of a strong coupling ($\gamma=0.06$). The frequencies which are circled correspond to the ones which are used to estimate the strength of the coupling in Sect. \ref{sect_discussion}.
\label{fig_ech_n}}
\end{figure}

The previous analogy relies on the fact that two modes only are affected during an avoided crossing: this corresponds to neglecting the coupling between the two considered modes and the other modes in the spectrum. Let us now push the analogy a step further, and consider the case where these other coupling terms play a significant role. We consider $n$ oscillators, instead of two only. We choose $n-1$ of them to simulate high-radial-order p modes, by giving them equidistant eigenfrequencies which are constant with $\lambda$. The last oscillator simulates a g mode. We have taken
\begin{eqnarray}
\omega_i(\lambda) & = & \omega_{i-1}(\lambda) + \Delta\omega, \;\;\; \hbox{for} \;\;\; i=2,\hdots,n-1 \\
\omega_n(\lambda) & = & \lambda
\end{eqnarray}
The numerical values of $\omega_1(\lambda)$ and $\Delta\omega$ (1 and 0.1 respectively) have been chosen to roughly match Fig. \ref{fig_evol_mm_intro}.

We then introduce a coupling $\alpha_i$ between the g mode and the $i\eme$ p mode. Normally the p modes should also be coupled to each other. However, since their eigenfrequencies remain equidistant at all times, we can neglect these coupling terms in our analogy. As $\lambda$ increases, the g mode will experience successive avoided crossings with all the p modes. According to Eq. \ref{eq_near_AC}, the deviation of the eigenfrequency very close to the $i\eme$ avoided crossing is $\delta\omega\sim\alpha_i/(2\omega_{i})$.
%, where $\omega_{0,i}$ is the frequency where $\omega_i(\lambda)=\omega_n(\lambda)$. 
To ensure that all the avoided crossings have a comparable intensity (as seems to be roughly the case in the models, see Fig. \ref{fig_evol_mm_intro}), we choose the different coupling terms $\alpha_i$ such that 
\begin{equation}
\alpha_i = \gamma \omega_{i}
\end{equation}
where $\gamma$ describes the strength of the coupling between the p-mode cavity and the g-mode cavity. We obtain the following system of equations
\begin{eqnarray}
\label{eqdiff_n}
\frac{\hbox{d}^2y_1(t)}{\hbox{d}t^2} & = & -\omega_1^2y_1+\gamma\omega_{1} y_n \\  \nonumber
& \vdots & \\ \nonumber
\frac{\hbox{d}^2y_{n-1}(t)}{\hbox{d}t^2} & = & -\omega_{n-1}^2y_{n-1}+\gamma\omega_{n-1} y_n \\ \nonumber
\frac{\hbox{d}^2y_n(t)}{\hbox{d}t^2} & = & -\omega_n^2y_n+\gamma\omega_1 y_1+\hdots+ \gamma\omega_{n-1} y_{n-1}\\ \nonumber
\end{eqnarray}

By writing the different oscillators $y_i(t)=c_i\exp(-i\omega t)$, the eigenfrequencies of the system are found by solving the eigenvalue problem $AC=\omega^2 C$ with
\begin{equation}
A=\begin{bmatrix}
\omega_1^2 & 0 & \cdots & 0 & -\gamma\omega_{1} \\
0 & \omega_2^2 & & \vdots & -\gamma\omega_{2} \\
\vdots & & \ddots & & \vdots \\
0 & \cdots &  & \omega_{n-1}^2 & -\gamma\omega_{n-1} \\
-\gamma\omega_{1} & \cdots & & -\gamma\omega_{n-1} & \omega_n^2 \vphantom{\vdots}
\end{bmatrix}
\end{equation}
and $C=\left[c_1, \cdots, c_n\right]$.

%\begin{figure}
%\includegraphics[width=8cm]{fig_cross_n.ps}
%\caption{Variations of the eigenfrequencies of $(n-1)$ $p$ modes coupled to a $g$ mode which undergoes avoided crossings with the $p$ modes (here, $n=10$).
%The dashed lines correspond to a "weak coupling" ($\alpha=0.05$), and the full lines to a "strong coupling" ($\alpha=0.35$).
%\label{fig_cross_n}}
%\end{figure}

The solution is plotted in Fig. \ref{fig_cross_n} for two different values of $\gamma$. For each value of the parameter $\lambda$, we can plot an \'echelle diagram with the eigenfrequencies of the coupled system (Fig. \ref{fig_ech_n}). In the case of a weak coupling, the ridge remains straight and the approximation that only two modes are affected by the avoided crossing is legitimate. However, for a strong coupling, more p modes in the neighborhood of the avoided crossing have a mixed behavior and the ridge is distorted as shown in Fig. \ref{fig_ech_n}.

%An interest of this analogy is that for each value of the parameter $\lambda$, we can plot an \'echelle diagram of the eigenfrequencies of the coupled system. In the case of uncoupled oscillators, the p modes are characterized by a straight ridge in an \'echelle diagram folded on the large separation (dotted line in Fig. \ref{fig_ech_n}). Fig. \ref{fig_ech_n} shows the effect of the coupling on the curvature of the ridge. In the case of a weak coupling, the approximation that only two modes are affected by the avoided crossing is legitimate and apart from these two modes, the ridge remains straight. However, as the coupling increases, more p modes in the neighborhood of the avoided crossing have a mixed behavior. Their frequencies are therefore modified and the curvature of the ridge is altered. Above the avoided crossing, the ridge is distorted to the right and below, it is distorted to the left. 

\subsection{Comparison with stellar models \label{sect_comp_models}}

\begin{figure*}
\begin{center}
\includegraphics[width=8cm]{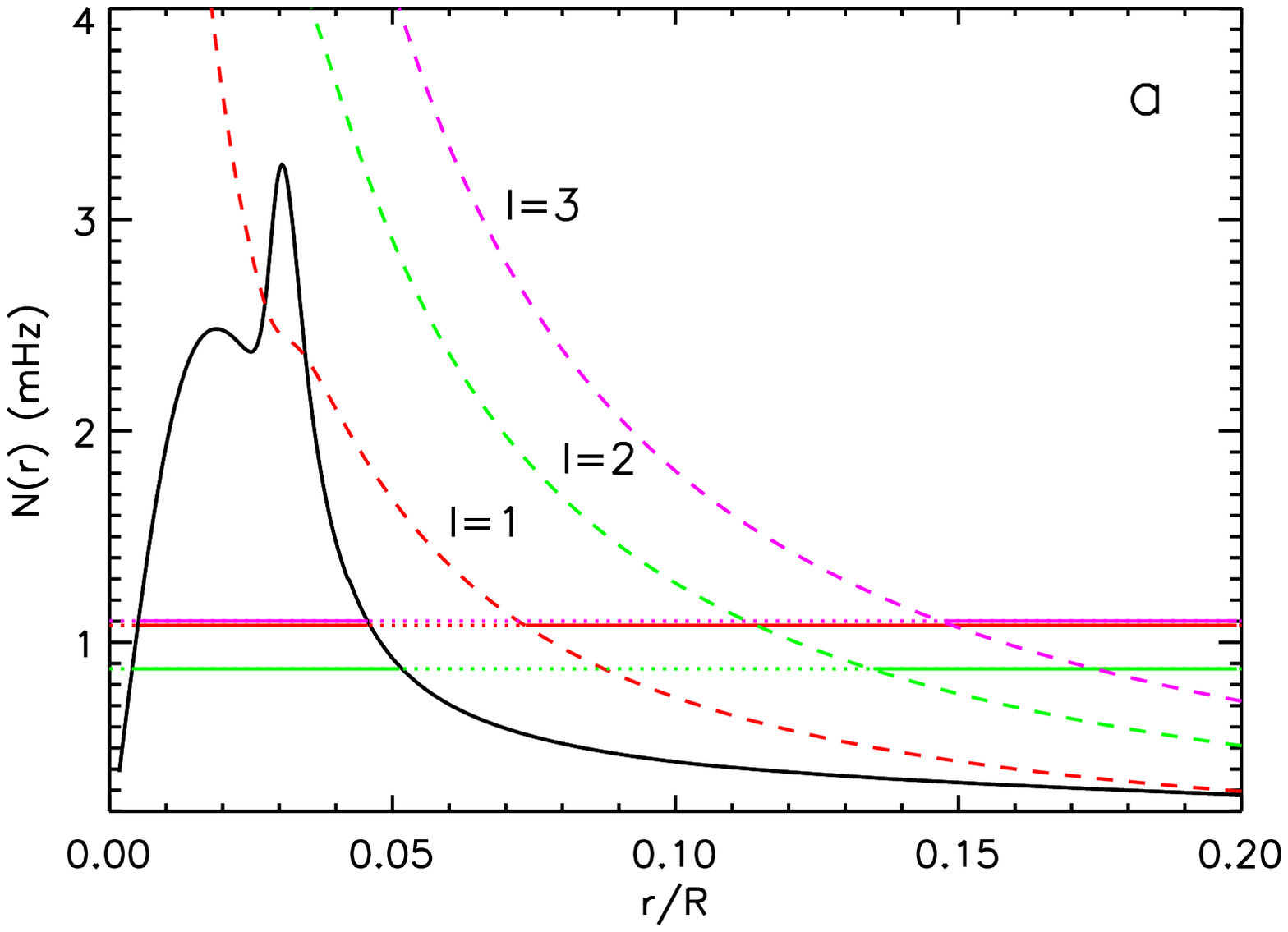}
\includegraphics[width=8cm]{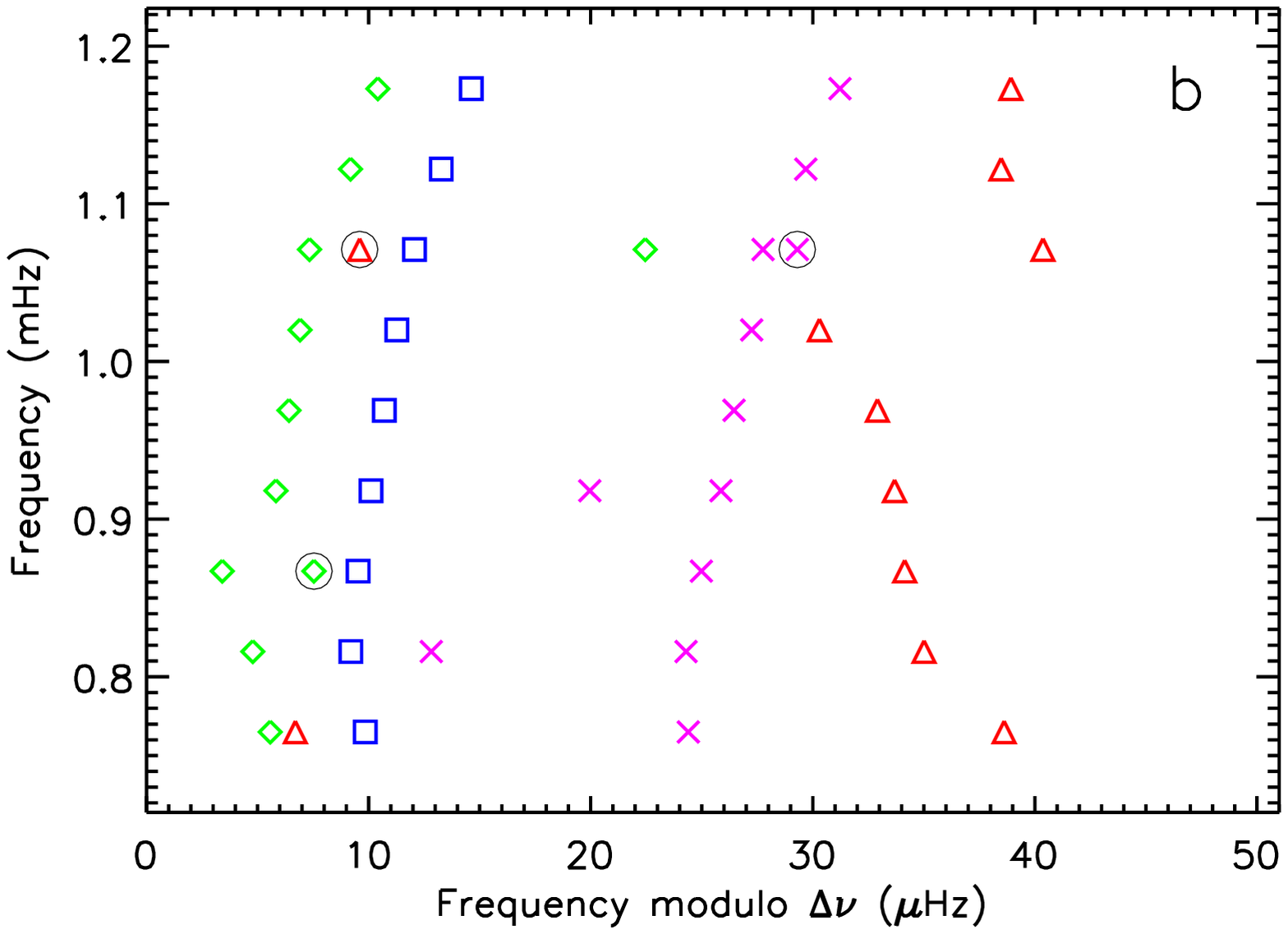}
\end{center}
\caption{\textit{Left}: Propagation diagram of a $1.3\,M_{\odot}$ post main sequence model. The \vaisala\ frequency is represented by the full black line and the Lamb frequencies by the dashed lines ($\ell=1$ red, $\ell=2$ green, $\ell=3$ purple). The propagation regions of three mixed modes of degrees $\ell=1,2$ and 3 are represented by the full colored lines and their evanescent regions by the dotted colored lines. \textit{Right}: Echelle diagram of the same model ($\ell=0$ blue squares, $\ell=1$ red triangles, $\ell=2$ green diamonds, $\ell=3$ purple crosses). The mixed modes represented in the left panel are circled.
%\'Echelle diagram of a $1.3\,M_{\odot}$ post main sequence model ($\ell=0$ blue squares, $\ell=1$ red triangles, $\ell=2$ green diamonds, $\ell=3$ purple crosses). \textit{Right}: Propagation diagram of the model. The \vaisala\ frequency is represented by the full black line and the Lamb frequencies by the dashed lines (same color code as left panel). The propagation regions of three mixed modes of degrees $\ell=1,2$ and 3 are represented by the full colored lines and their evanescent regions by the dotted colored lines.
\label{fig_ech_mod}}
\end{figure*}

%The analogy we presented shows that provided the coupling between the p-mode cavity and the g-mode cavity is strong enough, an avoided crossing of degree $\ell$ should induce a distortion in the ridge of degree $\ell$ in the \'echelle diagram. 

In a star, the coupling is due to the existence of an evanescent zone between the two cavities. The wider this zone, the weaker the coupling. Since the Lamb frequency increases with the dergee $\ell$, the coupling is stronger for modes of small degrees. We represented in Fig. \ref{fig_ech_mod}a the propagation diagram of a $1.3\,M_{\odot}$ model, evolved enough so that avoided crossings occur for low-degree modes (post main sequence). For this model, the ridges of the modes of degrees $1\leqslant\ell\leqslant3$ are shown in an \'echelle diagram in Fig. \ref{fig_ech_mod}b. Clearly, for modes of degree $\ell\geqslant2$, the currently adopted hypothesis that only two modes are affected by the avoided crossing is valid, the rest of the ridge being almost unaltered. However, for the $\ell=1$ avoided crossing, the coupling is strong enough so we observe a distortion in the ridge which is very comparable to the one we obtained in our simple analogy (bottom panel of Fig. \ref{fig_ech_n}). 

\subsection{Application to the case of \cible}

%We have shown in Sect. \ref{sect_analogy} that the $\ell=1$ avoided crossings induce a characteristic distortion in the $\ell=1$ ridge of the \'echelle diagram. We have therefore considered the possibility that the difference we observe in the curvature of the $\ell=1$ ridge compared to the $\ell=0$ ridge might be caused by an $\ell=1$ avoided crossing. 

By fine-tuning the mass and age of PoMS models, we managed to find a model with an $\ell=1$ avoided crossing in the low-frequency part of the frequency range of the observations (around $750\,\mu$Hz) and which also reproduces the observed mean value of the large separation (we will present in Sect. \ref{sect_nuac} a method to find these models in a systematic way). The \'echelle diagram of this model is represented in Fig. \ref{fig_ech_superpose}. We observe that the curvature of the $\ell=1$ ridge is far better reproduced by this model than by MS models. 
%Consequently, the profiles of the $\ell=1$ large separation and of the separation $\zeroun$ are in much better agreement with the observations. 
Interestingly, we note that by adjusting the age of the model, the frequency of the mode which behaves mainly as a g mode in the avoided crossing matches the frequency of the peak $\pi_1$ we mentioned before. 
%Finally, the small separation $\zerodeux$ is well reproduced by this model.

We conclude that the observed oscillation spectrum of \cible\ can be satisfactorily reproduced only if an $\ell=1$ avoided crossing causes a distortion the $\ell=1$ ridge in the \'echelle diagram. We thus obtain a firm detection of mixed modes in the spectrum of the star. We note that the avoided crossing occurs at the far bottom of the frequency range of the observations, which explains why we only detect the upper half of the ridge distortion described in Sect. \ref{sect_analogy}. We have also been able to identify the mode which has a mainly g-mode behavior in the avoided crossing (corresponding to the peak $\pi_1$). This will be very valuable when conducting a thorough modeling of the star, since this mode is the most sensitive to the center of the star. Finally, the detection of an $\ell=1$ avoided crossing in the spectrum of \cible\ enables us to establish that the star is in a PoMS stage.

%In the following two sections, we will try to determine the information about the structure of the core of \cible\ which are conveyed by the properties of the observed avoided crossing. 

\section{Stellar modeling using the avoided crossing: potential and strategy \label{sect_nuac}}

\subsection{Potential of avoided crossings with strong coupling \label{sect_potential}}

Mixed modes have a high potential in terms of seismic diagnostics since they are sensitive to the structure of the most central regions of the stars, which remains poorly understood. 

The frequencies of mixed modes depend on the profile of the \vaisala\ frequency in the stellar core. To understand the information which we can expect to derive from them, it is convenient to write the \vaisala\ frequency as a function of the following gradients:
\begin{equation}
\nabla\ind{ad} = \frac{\partial \ln T}{\partial \ln P} \bigg|_S \; , \; \;  \nabla = \frac{\partial \ln T}{\partial \ln P} \; , \; \;  \nabla\mu= \frac{\partial \ln \mu}{\partial \ln P},
\end{equation}
where $P$, $T$ and $\mu$ correspond to the stellar pressure, temperature and mean molecular weight, respectively. The subscript $S$ indicates that the definition is valid for constant entropy. The \vaisala\ frequency can be split in two different contributions:
\begin{equation}
N^2 = \frac{g}{H_P} \left( \nabla\ind{ad} - \nabla \right) + \frac{g}{H_P} \nabla\mu \label{eq_vaisala}
\end{equation}
where $g$ is the gravity inside the star and $H_P$ is the local pressure scale height. The first term in the right hand side of Eq. \ref{eq_vaisala} depends on the temperature stratification. The right term characterizes the dependance on the gradient of mean molecular weight $\nabla\mu$. These two components are shown in Fig. \ref{fig_diag_prop} for the model presented in Sect. \ref{sect_comp_models}.

%If the star was massive enough to have a convective core during the main sequence stage, a peak appears in $\nabla\mu$, which is the signature of the core withdrawal during the previous evolution of the star. The properties of this peak (location, width...) depend, among other things, on the mixing processes at the boundary between the convective core and the radiative zone. \\

%It has been known for long that the knowledge of the frequency of an avoided crossing could bring very valuable information about the inner structure of a star. Indeed, it provides an estimate of the frequency of the g mode which is involved. This opportunity is extremely precious, since the observation of pure g modes in main sequence stars is currently out of reach. 

Two different features of avoided crossings with strong coupling bring information about the structure in the center of the star:

\begin{enumerate}

\item \textit{The frequency at which the avoided crossing occurs} \\
%It has been known for long that the knowledge of the frequency of an avoided crossing could bring valuable information about the inner structure of a star. It indeed 
The knowledge of the frequency of an avoided crossing provides an estimate of the frequency of the g mode which is involved. This opportunity is extremely precious, since the observation of pure g modes in main sequence stars is currently out of reach. In the framework of the asymptotic theory (\citealt{tassoul80}), the frequency of a g mode of radial order $n$ and degree $\ell$ can be approximated by
\begin{equation}
\omega_{n,\ell} \sim \frac{\sqrt{\ell(\ell+1)}}{(n-1/2)\pi} \int_{r_1}^{r_2} \frac{N}{r} \hbox{d}r
\end{equation}
where $r_1$ and $r_2$ are the mode turning points in the g-mode cavity. In the following, we will consider avoided crossings involving low-radial-order g modes, for which the asymptotic approximation is no longer valid. However, their frequencies mainly depend on $\int_{r_1}^{r_2} N \hbox{d}r/r$. Having access to the frequency of a g mode gives information about the structure inside the g-mode cavity ($r_1\leqslant r\leqslant r_2$ in Fig. \ref{fig_diag_prop}). In particular, it is sensitive to the peak left in the profile of $\gradmu$ by the withdrawal of the convective core, for stars massive enough to have had one during the main sequence. This is the reason why the detection of mixed modes is expected as a means of constraining the amount of overshooting at the boundary of convective cores (e.g. \citealt{dziembowski91}).

\begin{figure}
\includegraphics[width=8.5cm]{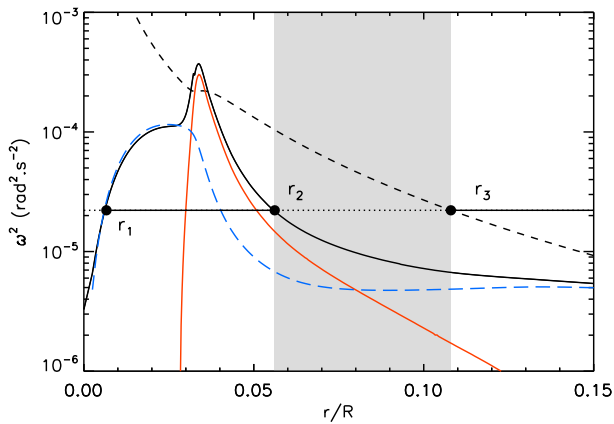}
\caption{Propagation diagram for a $1.3\,M_{\odot}$ post main sequence model which presents an $\ell=1$ avoided crossing. The \vaisala\ frequency (black solid line) is split in two contributions: the part linked to the temperature stratification (long-dashed blue line) and the one linked to the gradient of mean molecular weight (red solid line). The Lamb frequency for $\ell=1$ modes is represented by the black dashed line. The shaded area indicates the evanescent zone. The radii $r_1$ and $r_2$ correspond to the turning points of the g-mode cavity and $r_3$ the turning point of the p-mode cavity. 
\label{fig_diag_prop}}
\end{figure}

\vspace{0.1cm}
\item \textit{The intensity of ridge distortion} \\
We showed in Sect. \ref{sect_analogy} that more than two modes are affected by $\ell=1$ avoided crossings. The modification of their frequencies (characterized by a distortion of the $\ell=1$ ridge in the \'echelle diagram) depends on the strength of the coupling between the p-mode cavity and the g-mode cavity. This coupling mainly depends on the profile of the \vaisala\ frequency in the evanescent zone: the higher $N(r)$ in this region, the stronger the coupling. The intensity of the ridge distortion should therefore bring information about the structure of the star in the evanescent zone (i.e. in the region $r_2\leqslant r\leqslant r_3$ in Fig. \ref{fig_diag_prop}), complementary to those given by the frequency of the avoided crossing. We will illustrate this complementarity with the modeling of the star HD~49385 in Sect. \ref{sect_optim} and \ref{sect_discussion}.
%The distortion of the $\ell=1$ ridge due to an avoided crossing brings information on the strength of the coupling between the p-mode cavity and the g-mode cavity inside the star. The coupling mainly depends on the profile of the \vaisala\ frequency in the evanescent zone: the higher $N(r)$ in this region, the stronger the coupling. We therefore obtain information about the chemical composition and the temperature gradient in the region $r_2\leqslant r\leqslant r_3$ (see Fig. \ref{fig_diag_prop}), complementary to those given by the frequency of the avoided crossing. We will illustrate this complementarity with the modeling of the star HD~49385 in Sect. \ref{sect_nuac} and \ref{sect_optim}.

\end{enumerate}

\subsection{Limitations of traditional optimization techniques to model stars with avoided crossings}

As mentioned in Sect. \ref{sect_intro}, numerous theoretical works have been led about avoided crossings.
%and have shown that the frequency of mixed modes bears information about the structure of the inner parts of the star. 
However, very few studies have tried to find stellar models fitting the properties of an avoided crossing. 
The first reason for this is that until very recently, the spectra in which the detection of mixed modes was claimed were too imprecise to lead such an investigation.
%no avoided crossing had been detected in the oscillation spectra of observed stars. 
%Ground observations of $\eta$ Boo have shown the probable presence of an avoided crossing (a developper, \citealt{kjeldsen95b}, \citealt{guenther96}, \citealt{dimauro03}). 
The second reason is that it presents some inherent difficulties linked to the fact that the avoided crossings occur on a very short timescale (typically of the order of 1 Myr or less) compared to the stellar evolution timescale. The usual approach to model a star consists in computing a grid of models with varying stellar parameters (mass, age, etc.) to find the optimal model which minimizes the $\chi^2$ function defined in Eq. \ref{eq_chi2}. Applying this procedure with a time step of the order of the avoided crossing timescale is not feasible, since it would require to compute a tremendous number of models. On the other hand, with a larger time step, we have a very low probability to find models which correctly reproduce the frequency of the avoided crossing and we therefore miss the best-matching models. We note that it would also be impractical to bluntly apply an automatic minimization using for instance the algorithm of Levenberg-Marquardt, as prescribed by \cite{miglio05}. Indeed, this technique is based on the computation of the Hessian matrix at each iteration and it thus requires to input a step for the time derivatives. The gap between the stellar evolution timescale and that of the avoided crossing makes it impossible to find a satisfactory time step for the procedure.

%\textbf{Remarque sur la courbure de la cr\^ete : si on n'a peu ou pas de mod\`eles qui reproduisent $\nuac$, on ne peut pas d\'eduire d'info sur le couplage et donc sur la structure dans la zone \'evanescente.}

\begin{figure}
\includegraphics[width=8.5cm]{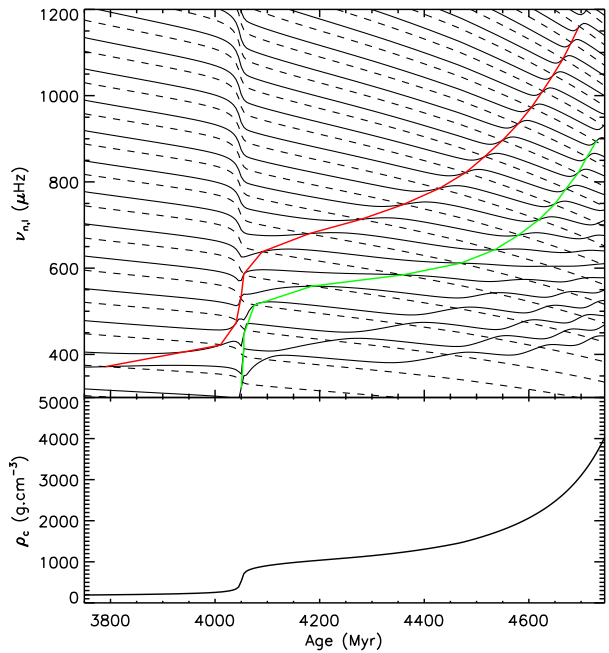}
% En fait il s'agit d'un modele de 1.28 Msun.
\caption{\textit{Top:} Evolution of the frequencies of $\ell=0$ modes (dashed black lines) and $\ell=1$ modes (full black lines) with age for a $1.3\,M_{\odot}$ model. The frequencies of the first two $\ell=1$ uncoupled g modes are also represented (g$_1$ mode in red and g$_2$ mode in green). \textit{Bottom:} Evolution of the central density $\rho\ind{c}$ with age for the same model.
\label{fig_evol_mm}}
\end{figure}

%First of all we need to find a way of following an avoided crossing during the evolution of the studied models. It is not that simple since the g modes experience a series of consecutive avoided crossings as the star evolves. For instance in the example given in Sect. \ref{sect_}, at an age of ?? Myr, the mode $\tilde{n}=14$ is mainly trapped in the core and behaves as a g$_1$ mode. ?? Myr later, it is the mode $\tilde{n}=15$ which behaves as the g$_1$ mode. The best way to follow the avoided crossing would to have access to the frequency of the g$_1$ mode in the case where the g-mode cavity would not be coupled to the p-mode cavity. 

%It is of course best described by the frequency of the g mode it involves in the case where the g-mode cavity would not be coupled to the p-mode cavity. Since the cavities are actually coupled in the models we do not have access to this frequency.

%The evolution of an avoided crossing is best described by the frequency of the g mode which it involves. It indeed varies much faster than the frequencies of the p modes. However, as can be seen in Fig. \ref{fig_evol_mm}, this mode experiences a series of avoided crossings as the star evolves and 

%For instance, at an age of about 4500 Myr, the mode p$_{11}$ behaves as a g$_{1}$ mode, whereas at an age of 4650 Myr, it is the mode p$_{13}$ which presents this behavior. 

%for which $n=11$ behaves mainly as a g mode

\subsection{Narrowing down the dimension of the model space}

We now try to adapt the traditional grid-of-model approach to the special case of targets for which mixed modes are detected. For this purpose, we show that we can overcome the obstacles we just mentioned by no longer considering the stellar age and mass as free parameters of our fit. 
%We start by investigating the way the frequency of an avoided crossing varies with the different parameters of the models.

\subsubsection{Relation between $\nuac$ and the stellar age \label{sect_nuac_age}}

\begin{figure*}
\begin{center}
\includegraphics[width=8cm]{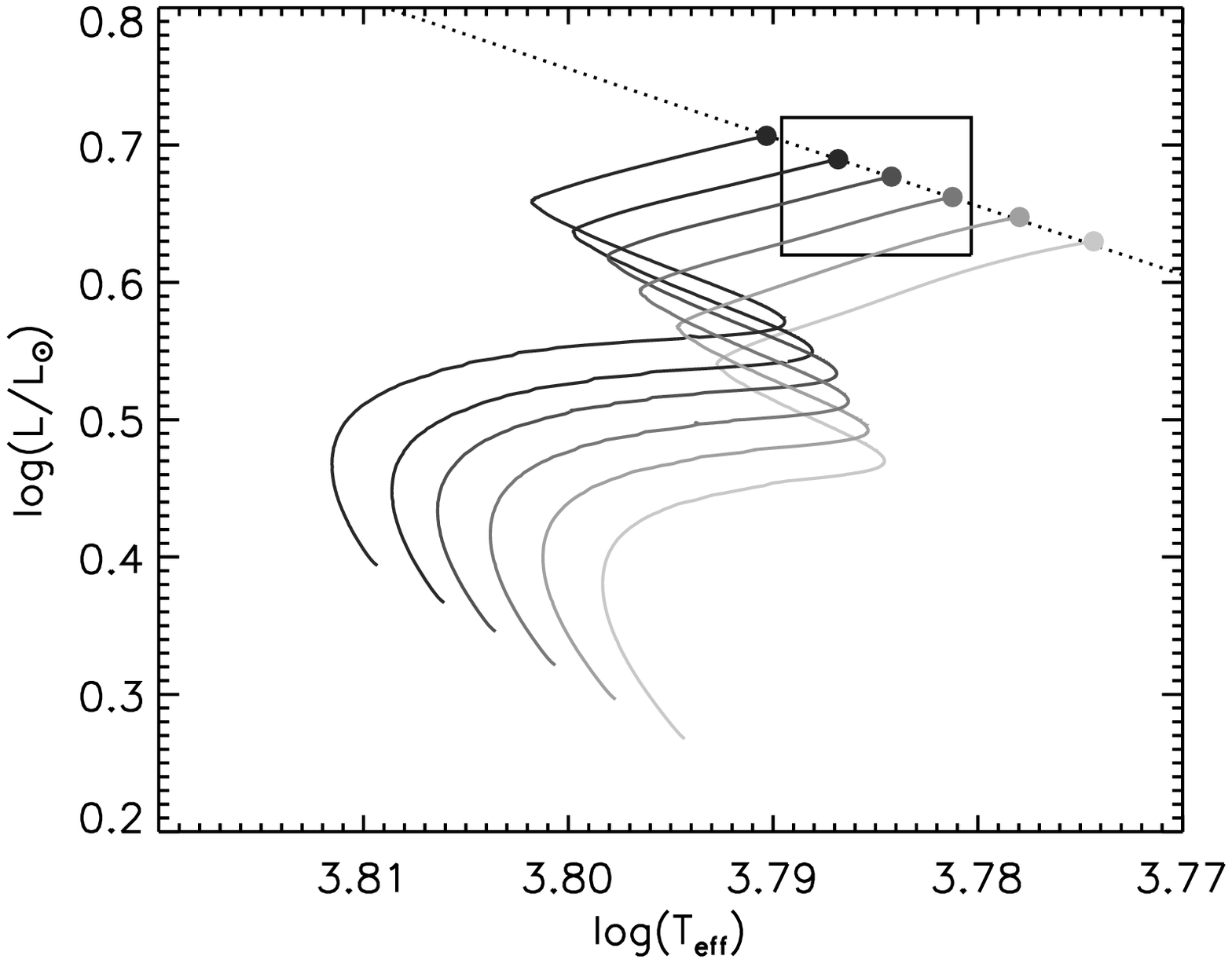}
\includegraphics[width=8cm]{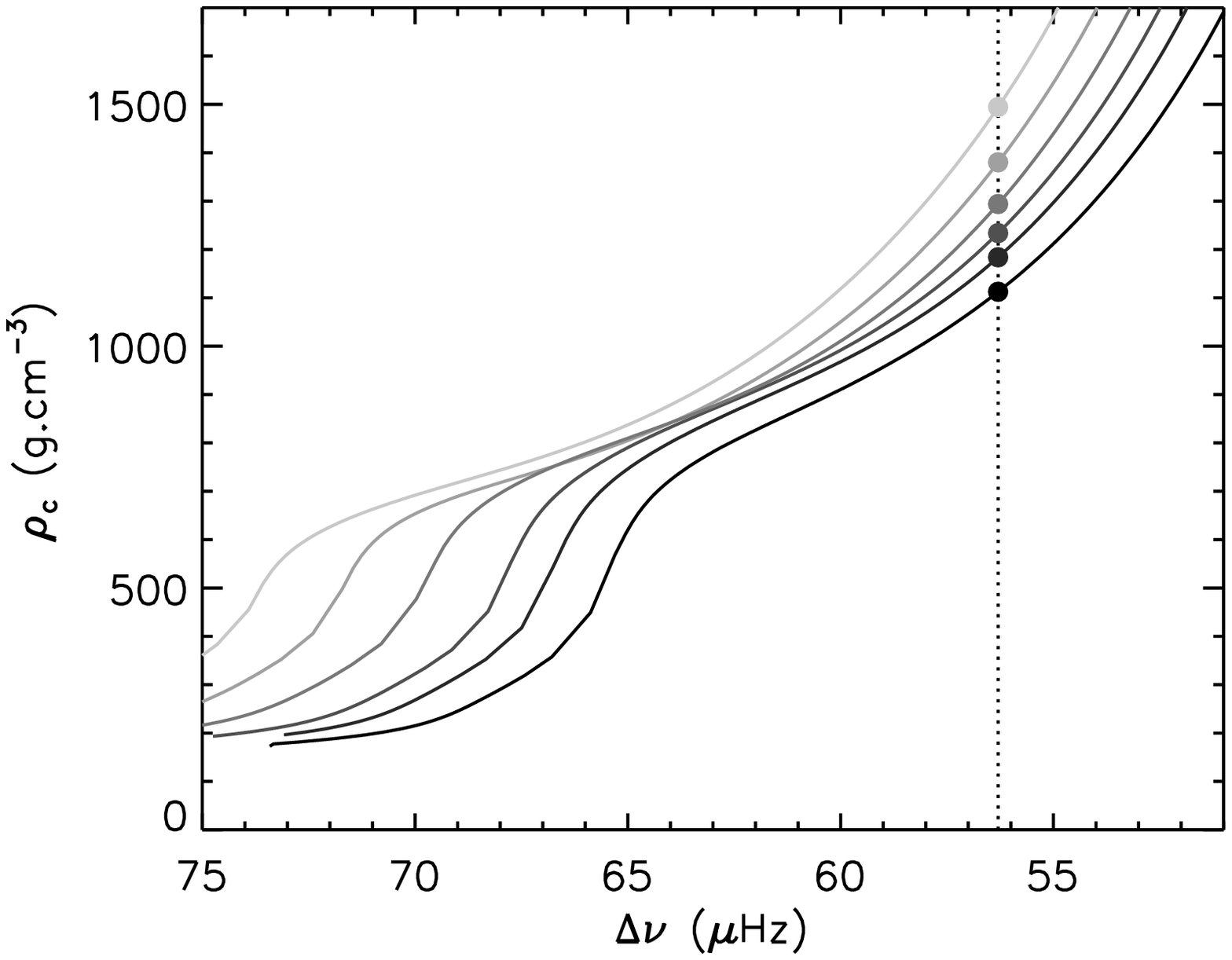}
\end{center}
\caption{\textit{Left}: Evolutionary tracks in the HR diagram for models of masses ranging from $1.23\,M_{\odot}$ (lightest symbols) to $1.31\,M_{\odot}$ (darkest symbols), with given physics. The filled circles indicate the points where the models reach $\lsmean=\lsmean\ind{obs}$ for \cible\ (this location is materialized by the dotted line). \textit{Right}: Evolution of the central density for the same models. The value of the mean large separation has been used as an indicator of the age, for more convenience.
\label{fig_hr_mass}}
\end{figure*}

The frequency at which an avoided crossing occurs (further noted $\nuac$) corresponds to the frequency of the uncoupled g mode it involves and is therefore linked to the profile of the \vaisala\ frequency in the core (see Sect. \ref{sect_potential}). Fig. \ref{fig_evol_mm} shows the variations of the frequencies of the first two uncoupled g modes during the evolution of a $1.3\,M_{\odot}$ model (the sharp variations of the mode eigenfrequencies around the age of 4050 Gyr in this plot correspond to the transition between the MS stage and the PoMS stage).

During the main sequence, the first term in the right hand side of Eq. \ref{eq_vaisala} nearly vanishes in the center since the temperature gradient is almost adiabatic in the convective core. The g-mode cavity essentially corresponds to the chemically inhomogeneous zone left by the withdrawal of the core. At this point, the \vaisala\ frequency is still low, so that low-degree g modes do not reach the frequency range of high-order p modes and no avoided crossing can be observed.

At the end of the main sequence, when the hydrogen reserves are exhausted in the core, the star is left with an almost isothermal helium core. The inner regions contract until nuclear reactions are triggered in the layers where hydrogen remains. The evolution of the \vaisala\ frequency is then mainly determined by the value of the central density $\rho\ind{c}$. To explain this, we remark from Eq. \ref{eq_vaisala} that $N$ depends on the factor $g/H_P$, which can also be written $\rho g^2/P$.
%(if we avoid the fast contraction period that follows the end of the main sequence and suppose that the star is in hydrostatic equilibrium)
In the most central regions, the gravity term can be approximated as $g(r)\sim G\rho\ind{c}r$. In that case, we have
\begin{equation}
\frac{g}{H_P} \, \sim \, \frac{\rho\ind{c}^3}{P\ind{c}} r^2 \, \sim \, \frac{\rho\ind{c}^2 \mu\ind{c} }{T\ind{c}} r^2 \label{eq_BV_rhoc}
\end{equation}
where $P\ind{c}$ and $T\ind{c}$ are the central pressure and temperature. The models show that the central temperature hardly changes in the isothermal core during the evolution in the post main sequence phase. The mean molecular weight is maintained constant since the chemical composition in the core does not vary. On the contrary, the contraction of the star causes the central density to increase by more than one order of magnitude compared to its value at the end of the MS, as can be seen in the lowest panel of Fig. \ref{fig_evol_mm}. As a result, the factor $g/H_P$ and thus the \vaisala\ frequency mainly depend on the evolution of the central density $\rho\ind{c}$. This is confirmed by Fig. \ref{fig_evol_mm} which shows the tight relation between the evolution of $\nuac$ and $\rho\ind{c}$ for a $1.3\,M_{\odot}$ model. 

We conclude that the avoided crossing frequencies monotonically increase during the evolution of the star. We can therefore find for every model (with fixed mass and physics) one stellar age such that the avoided crossing occurs at the same frequency as in the observations. For the model represented in Fig. \ref{fig_evol_mm}, an age of $4374\pm0.8$ Myr satisfies this condition (the error bar is determined by requiring a fit of $\nuac\expon{obs}$ within a 1-$\sigma$ error bar). The precision we obtain is such that we can consider the age to be entirely determined by the value of $\nuac$ (for models at given mass and physics) and we can cease to consider the age as a free parameter of the fit. This is very interesting since it solves the problem of the choice of a time step for the grid of models. 

We note that by using this method, we overlook the models which do not fit the frequency of the avoided crossing within 1-$\sigma$ error bars. However, since avoided crossings occur on a small timescale compared to the stellar evolution timescale, modifying $\nuac$ in a model will result in almost no change in the other stellar parameters. It will only induce an increase in the contribution of the term including $\nuac$ in the $\chi^2$ defined in Eq. \ref{eq_chi2} with no significant change in the other contributions. Overlooking these models should therefore not modify the optimal model we obtain nor the error bars we derive from it.

\subsubsection{Correspondence between $(\Delta\nu,\nuac)$ and $(M,\hbox{age})$ \label{sect_nuac_mass}}

\begin{figure}
\begin{center}
\includegraphics[width=8cm]{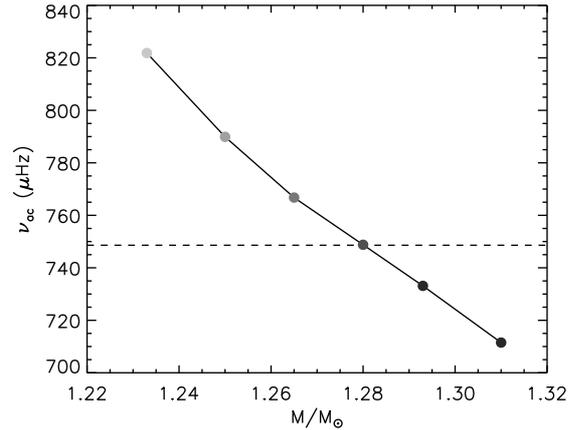}
\end{center}
\caption{Frequency of the avoided crossing for the models of Fig. \ref{fig_hr_mass} which reproduce the observed mean large separation of \cible. The dashed line indicates the frequency of the observed avoided crossing $\nuac\expon{obs}$ for \cible.
\label{fig_nuac_mass}}
\end{figure}

We just saw that for a model with given mass and physics, we can find an age such that $\nuac = \nuac\expon{obs}$. However, it is important to notice that the mean large separation of the star monotonically decreases with the age since the stellar radius keeps increasing (even during the sharp transition between the main sequence and the post main sequence stage). For every model, there exists an age at which the observed value of the mean large separation is reproduced. This age is a priori different from the one for which the frequency of the avoided crossing is reproduced. So which are the models that fit both conditions simultaneously? We will further refer to this joined condition as condition $\mathcal{C}$, i.e.
\begin{equation}
\mathcal{C} \; : \; 
\begin{cases}
\lsmean\ind{mod} = \lsmean\ind{obs} \\
\nuac\expon{mod} = \nuac\expon{obs}
\end{cases}
\end{equation}

To answer this question, we search among the models fitting the observed value of the mean large separation those which also reproduce the frequency of the avoided crossing. As explained in Appendix \ref{app_dn}, the 
iso-$\Delta\nu$ region in the HR diagram roughly corresponds to a line of slope 5 ($L\propto T\ind{eff}^5$).
%for models satisfying condition $\mathcal{C}$ among models reproducing the observed value of the mean large separation. As is explained in Appendix \ref{app_dn}, these models are roughly located along a line of iso-$\Delta\nu$ in the HR diagram.
%, which varies relatively little as a function of the stellar parameters. We give a brief explanation to this phenomenon in the Appendix \ref{app_dn}. 
Fig. \ref{fig_hr_mass} shows the evolutionary tracks of several models with the same given physics and different masses. We observe that as the stellar mass increases, the line of the Terminal Age Main Sequence (TAMS) gets closer to the line of iso-$\Delta\nu$. Consequently, the more massive stars are closer to the TAMS when they reach the observed value of the large separation. This means that they have smaller central density, as can be seen in Fig. \ref{fig_hr_mass} (where we found it convenient to use the mean large separation instead of the age as an indicator of the evolution). Since we mentioned that the evolution of $\nuac$ is determined by that of $\rho\ind{c}$, we conclude that $\nuac$ should decrease with increasing mass. This is confirmed by Fig. \ref{fig_nuac_mass}. This result should be kept in mind, as it will frequently be used in the following.
%the bigger the mass, the closer to the TAMS the star is when it reaches the observed value of the large separation. As a result, the more massive stars 
%We have seen that the frequency of the avoided crossing increases as the star moves away from the TAMS. It is therefore logical that $\nuac$ should decrease with increasing mass, which is confirmed by Fig. \ref{fig_nuac_mass}. 

We reach the conclusion that there exists \textit{one and only one} value of the stellar mass and age for which both the mean value of the large separation and the frequency of the avoided crossing are correctly fitted (for instance with the physics used in Fig. \ref{fig_hr_mass} and \ref{fig_nuac_mass}, it corresponds to a mass of about $1.28\,M_{\odot}$). We will further note $\tilde{M}$ and $\tilde{\tau}$ the stellar mass and age which satisfy condition $\mathcal{C}$. It is striking to remark that for a given physics, the two seismic parameters $\lsmean$ and $\nuac$ alone provide an estimate of the stellar mass and age, without having to resort to the use of any classical constraints or other seismic parameters. This estimate is also extremely precise. For example, if we fix all stellar parameters other than the mass and age in the case shown in Fig. \ref{fig_nuac_mass}, the uncertainty we obtain on the mass is as low as $4\times10^{-4}\,M_{\odot}$. We therefore propose to narrow down the dimension of the model space by eliminating the mass and age from the set of free parameters, these two quantities being determined by $(\lsmean,\nuac)$. Of course, the values of $\tilde{M}$ and $\tilde{\tau}$ depend on the physics we use to model the star. Their dependance with the different stellar parameters will be studied in Sect. \ref{sect_M_tilde}, after describing the method we adopted in this study to find the models which satisfy condition $\mathcal{C}$.

%It is interesting to note that the two seismic parameters $\lsmean$ and $\nuac$ alone provide an estimate of the stellar mass and age, without having to resort to the use of any classical constraints or other seismic parameters. This estimate is also extremely precise. For example, if we fix all stellar parameters other than the mass and age in the case shown in Fig. \ref{fig_nuac_mass}, the uncertainty we obtain on the mass is as low as $4\,10^{-4}\,M_{\odot}$. Of course, this is true only at given physics. The dependance of the mass $\tilde{M}$ with different other stellar parameters will be studied in Sect. \ref{sect_nuac_conv} to \ref{sect_nuac_ov}, after describing the method we adopted in this study to find the models which verify condition $\mathcal{C}$.

\subsection{Searching for models satisfying condition $\mathcal{C}$ \label{sect_searchC}}

We now briefly describe the procedure we followed to obtain the values of $\tilde{M}$ and $\tilde{\tau}$ for each set of free parameters.

As mentioned in Sect. \ref{sect_nuac_age}, we consider the stellar age to be fixed by imposing $\nuac=\nuac\expon{obs}$. Until now, we have assumed that the frequency of the avoided crossing corresponds to the frequency of the uncoupled g mode it involves. Of course, when computing the eigenfrequencies of a model, we do not have access to this quantity. We thus chose to assimilate the frequency of the avoided crossing to the frequency of the mode which behaves mainly as a g mode. In the oscillation spectrum of \cible, it corresponds to the peak labeled as $\pi_1$, whose frequency has been found to be $\nu_{\pi_1} = 748.6 \pm 0.23 \, \mu\hbox{Hz}$ by D10, and in the models it is the mode $\nu_{1,11}$. \textit{For each computed model}, the age is determined so that we have $\nu_{1,11}=\nu_{\pi_1}$. Special care has to be taken in this procedure since the evolution of the $\ell=1$ mode frequencies with the age is not monotonic due to the avoided crossings, as can be clearly seen in Fig. \ref{fig_evol_mm}. As a result, there are several ages (three) for which $(\nu_{1,11}=\nu_{\pi_1})$ and only one among them corresponds to the one we search for.

We need to determine the stellar mass for which the models fitting the frequency of the avoided crossing also reproduce the observed large separation. This mass is the one that minimizes the merit function
\begin{equation}
\chi^2_{\lsmean} \equiv \frac{(\lsmean\ind{mod}-\lsmean\ind{obs})^2}{\sigma_{\lsmean}^2}
\end{equation}
where $\lsmean\ind{obs}$ is the mean large separation of the observed radial modes, $\sigma_{\lsmean}$ the 1-$\sigma$ error-bar and $\lsmean\ind{mod}$ the corresponding quantity for the models (after correcting for the surface effects as described in Sect. \ref{sect_comp_mod_obs}). Only the radial modes are considered here since the large separations of non-radial modes are affected by avoided crossings. We remark that for models reproducing $\nuac\expon{obs}$, the mean large separation of radial modes varies almost linearly with the stellar mass. The function $\chi^2_{\lsmean}$ is therefore almost quadratic and the method of Newton is particularly efficient to determine the optimal mass $\tilde{M}$ (obtained in only a couple of iterations).

\subsection{Dependance of $\tilde{M}$ on the stellar parameters \label{sect_M_tilde}}

\begin{figure*}
\begin{center}
\includegraphics[width=6cm]{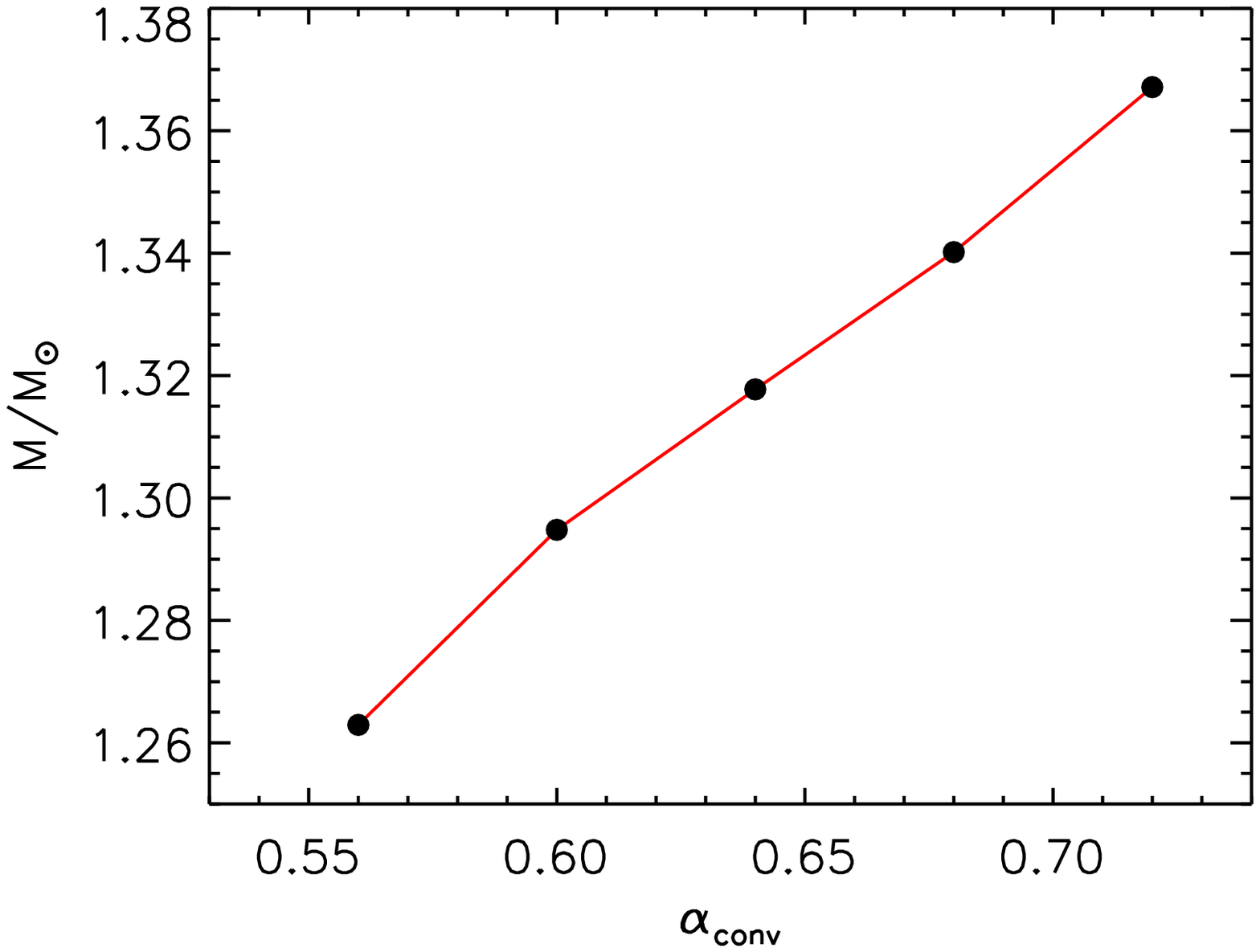}
\includegraphics[width=6cm]{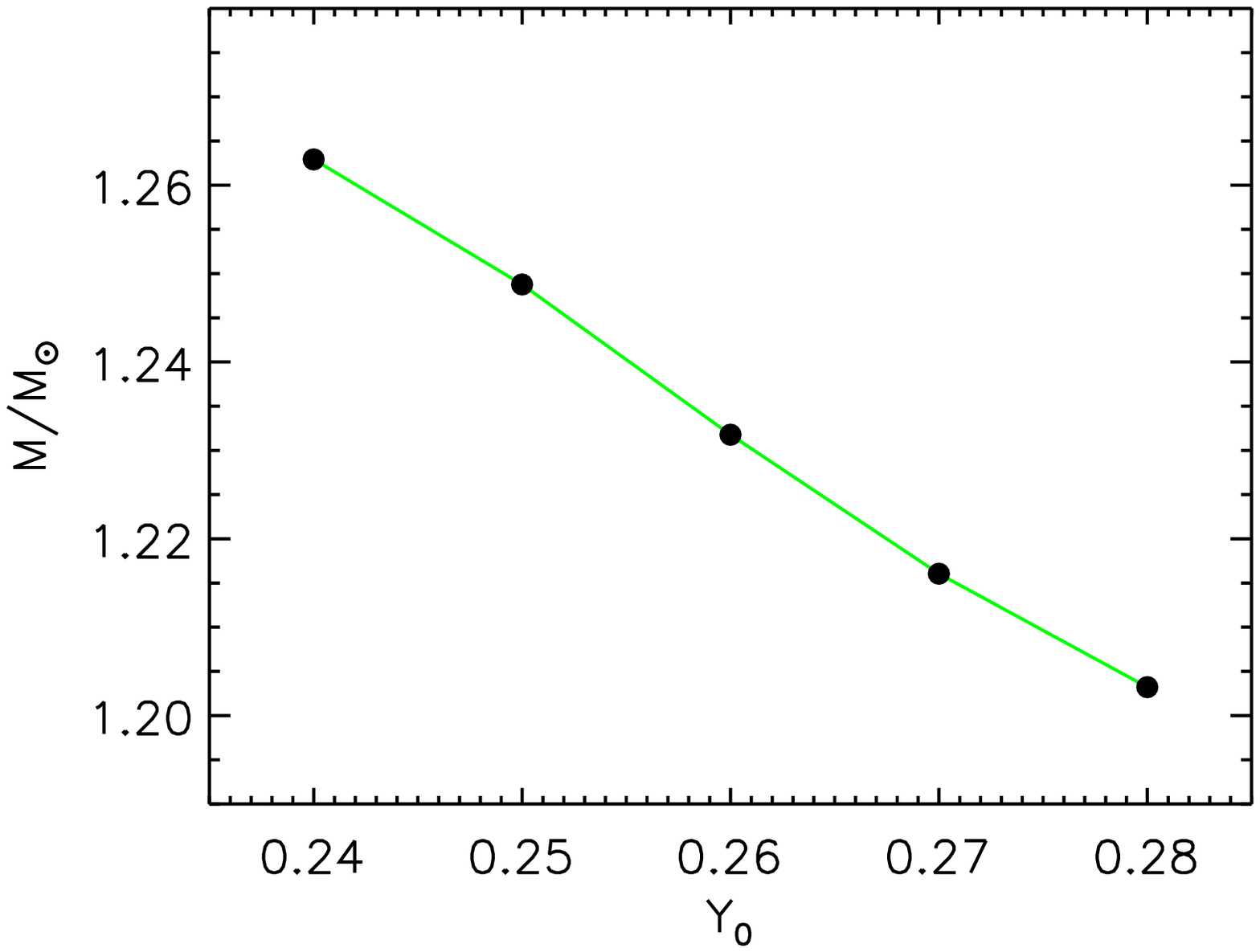}
\includegraphics[width=6cm]{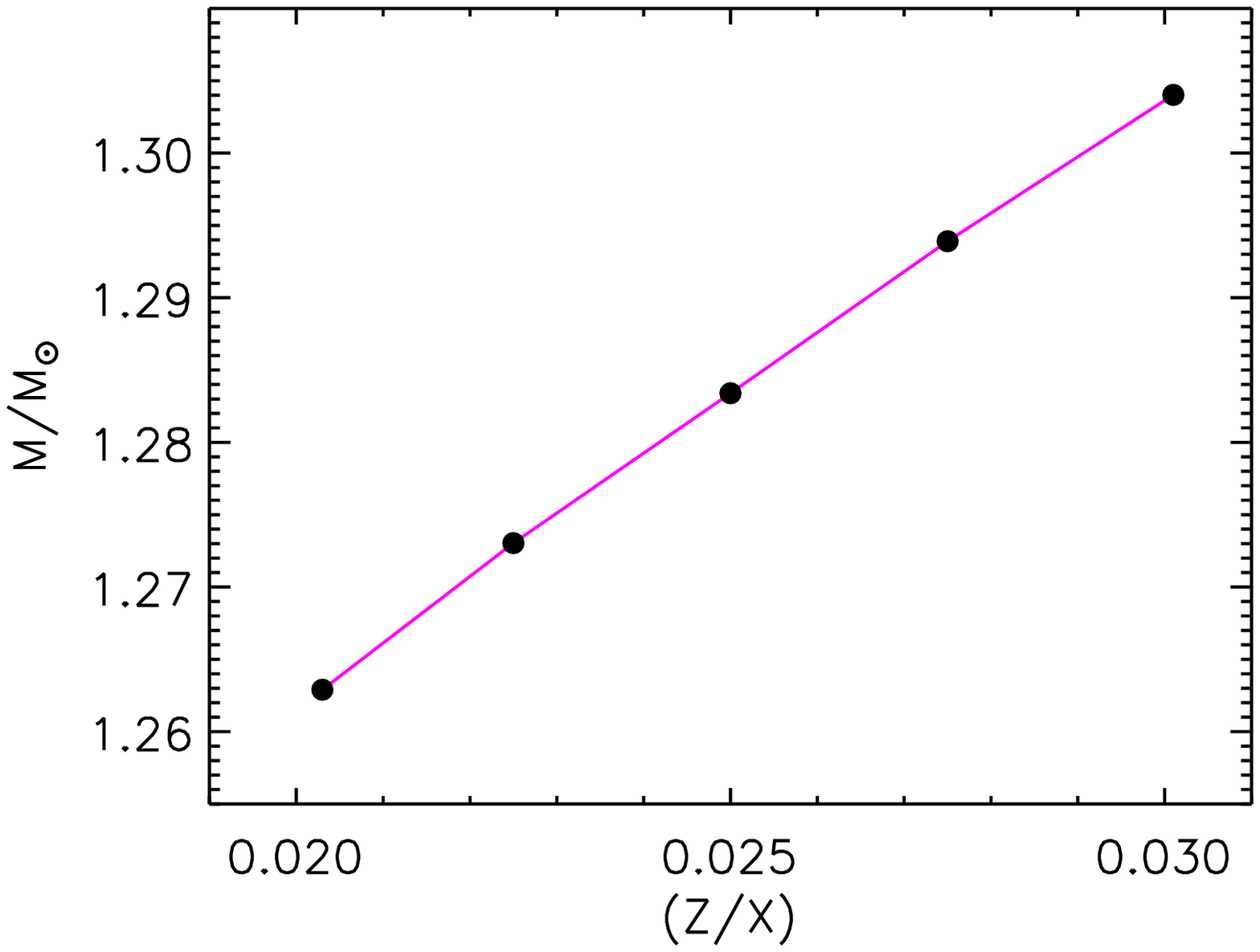}
\end{center}
\caption{Variations of the mass $\tilde{M}$ when changing the mixing length parameter (\textit{left}), the initial helium abundance (\textit{middle}) and the metallicity (\textit{right}) from model $S_0$ (whose properties are given in Table \ref{tab_models}).
%\textbf{Top}: Variations of the mass $\tilde{M}$ when changing the mixing length parameter in model $S_0$ whose properties are given in Table \ref{tab_models} (\textbf{left}). Evolutionary tracks in the HR diagram of models $S_0$, $S_{1a}$ and $S_{1b}$. The dotted line indicates the location where $\lsmean=\lsmean\ind{obs}$.
%Effect of a change of the mixing length parameter $\alpha\ind{conv}$ in model $\mathcal{M}_0$ (see text) on the mass $\tilde{M}$ which verifies condition $\mathcal{C}$. \textbf{Right}: Evolutionary tracks in the HR diagram of model $\mathcal{M}_0$ (mass $\tilde{M}_0$ and $\alpha\ind{conv}=0.56$, blue line), a model computed with the same mass but $\alpha\ind{conv}=0.60$ (red dashed line) and 
%Dashed lines (resp. full lines) represent the evolutionary tracks of models for which $\alpha\ind{conv}=0.56$ (resp. $\alpha\ind{conv}=0.60$). 
\label{fig_nuac_conv}}
\end{figure*}

By applying the procedure we just described, the mass $\tilde{M}$ and age $\tilde{\tau}$ can be determined for any given set of free parameters. This gives us the opportunity to study the dependance of $\tilde{M}$ with the different stellar parameters which are varied in this study (the mixing length parameter $\alpha\ind{conv}$, the initial helium abundance $Y_0$, the metallicity $[Z/X]$ and the amount of core overshooting $\aov$). This will provide us with helpful insights when discussing the results of our optimization in Sect. \ref{sect_discussion}. This should also give some hints of how the mass should be varied when studying other stars with avoided crossings.

We start from a model $S_0$ with a given set of stellar parameters (see Table \ref{tab_models}). From what has just been said, there exists one stellar mass $\tilde{M}_0$ and age $\tilde{\tau}_0$ which verifiy condition $\mathcal{C}$ with these parameters. By varying the stellar parameters one by one we try to understand the variations of the optimal mass $\tilde{M}$ with each of them.

%In the following sections, we study the dependance of the mass $\tilde{M}$ on the different stellar parameters. This will provide helpful insights when searching for optimal models in Sect. \ref{sect_optim}. We start from a model $\mathcal{M}_0$ with a given set of stellar parameter (in Fig. \ref{fig_nuac_conv}, we chose $\alpha\ind{conv}=0.56$, $Y_0=0.24$, $[Z/X]=0.0203$ and $\alpha\ind{ov}=0$ with the abundances of AGS05). From what has just been said, there exists one stellar mass $\tilde{M}_0$ and age $\tilde{\tau}_0$ which verifiy condition $\mathcal{C}$ with these parameters. By varying the stellar parameters one by one we try to understand the variations of the optimal mass $\tilde{M}$ with each of them.

\begin{figure}
\begin{center}
\includegraphics[width=8.5cm]{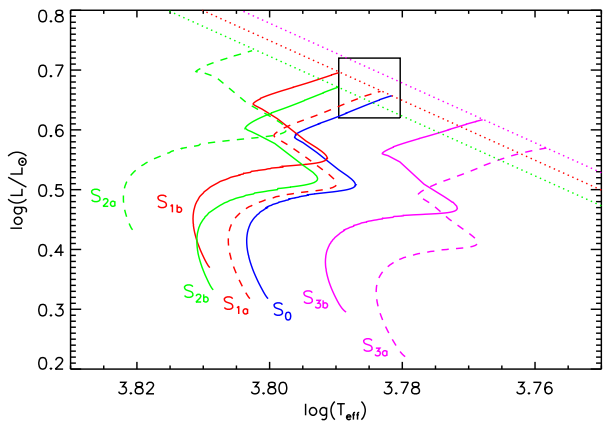}
\end{center}
\caption{Evolutionary tracks in the HR diagram of models $S_0$, $S\ind{1a}$, $S\ind{1b}$, $S\ind{2a}$, $S\ind{2b}$, $S\ind{3a}$ and $S\ind{3b}$ (see text and Table \ref{tab_models}).
\label{fig_hr_conv}}
\end{figure}

\begin{table}
  \centering
  \caption{Parameters of the models used to determine the variations of the mass $\tilde{M}$ (which verifies condition $\mathcal{C}$) with the stellar parameters.
  \label{tab_models}}
\begin{tabular}{c c c c c c}
\hline \hline
\T \B Model & $\alpha\ind{conv}$ & $Y_0$ & Mixture & Mass \\
\hline 
\T \B $S_0$ & 0.56 & 0.24 & AGS05 & $\tilde{M}_0$ ($1.263\;M_{\odot}$) \\
\hline
\T $S\ind{1a}$ & 0.60 & - & - & $\tilde{M}_0$ \\
\B $S\ind{1b}$ & 0.60 & - & - & $\tilde{M}_1$ ($1.295\;M_{\odot}$) \\
\hline
\T $S\ind{2a}$ & - & 0.28 & - & $\tilde{M}_0$ \\
\B $S\ind{2b}$ & - & 0.28 & - & $\tilde{M}_2$ ($1.202\;M_{\odot}$) \\
\hline
\T $S\ind{3a}$ & - & - & GN93 & $\tilde{M}_0$ \\
\B $S\ind{3b}$ & - & - & GN93 & $\tilde{M}_3$ ($1.308\;M_{\odot}$) \\
\hline
\end{tabular}
\end{table}

\subsubsection{Dependance of $\tilde{M}$ on the mixing length parameter \label{sect_nuac_conv}}

%Let us start from a given mixing length parameter $\alpha_0$. From what has just been said, there exists one stellar mass $\tilde{M}_0$ and age $\tilde{\tau}_0$ which verifiy condition $\mathcal{C}$. We here study the impact on the optimal mass $\tilde{M}$ of a modification of the mixing length parameter to a value of $\alpha_1>\alpha_0$. 

We modified the mixing length parameter of model $S_0$ without changing the other parameters and we determined the new mass $\tilde{M}$ by applying the procedure described in Sect. \ref{sect_searchC}. Fig. \ref{fig_nuac_conv}a shows that $\tilde{M}$ linearly increases with $\alpha\ind{conv}$.

The increase of $\tilde{M}$ with $\alpha\ind{conv}$ can be understood as follows. Let us modify the mixing length parameter from the value $\alpha_0$ of model $S_0$ to a value $\alpha_1>\alpha_0$. If we keep the same mass $\tilde{M}_0$ (model $S\ind{1a}$ in Table \ref{tab_models}), an increase of the mixing length induces an increase of the convective flux in the envelope and therefore causes the star to contract slightly. Since the stellar luminosity remains unchanged, the decrease of the radius is compensated by an increase of the effective temperature. The evolutionary track is thus horizontally translated to the left in the HR diagram when the mixing length goes from $\alpha_0$ to $\alpha_1$ at constant mass (see Fig. \ref{fig_nuac_conv}b). As a result, for $\alpha\ind{conv}=\alpha_1$, the star is further away from the TAMS when it reaches the observed value of the large separation and the frequency of the avoided crossing is too high compared to the observations. Based on what was said in Sect. \ref{sect_nuac_mass}, we need to increase the stellar mass to restore the agreement on the avoided crossing frequency while keeping the agreement on the large separation (model $S\ind{1b}$). We therefore have $\tilde{M}_1>\tilde{M}_0$.

\subsubsection{Dependance of $\tilde{M}$ on the helium abundance \label{sect_nuac_y0}}

\begin{figure*}
\begin{center}
\includegraphics[width=8cm]{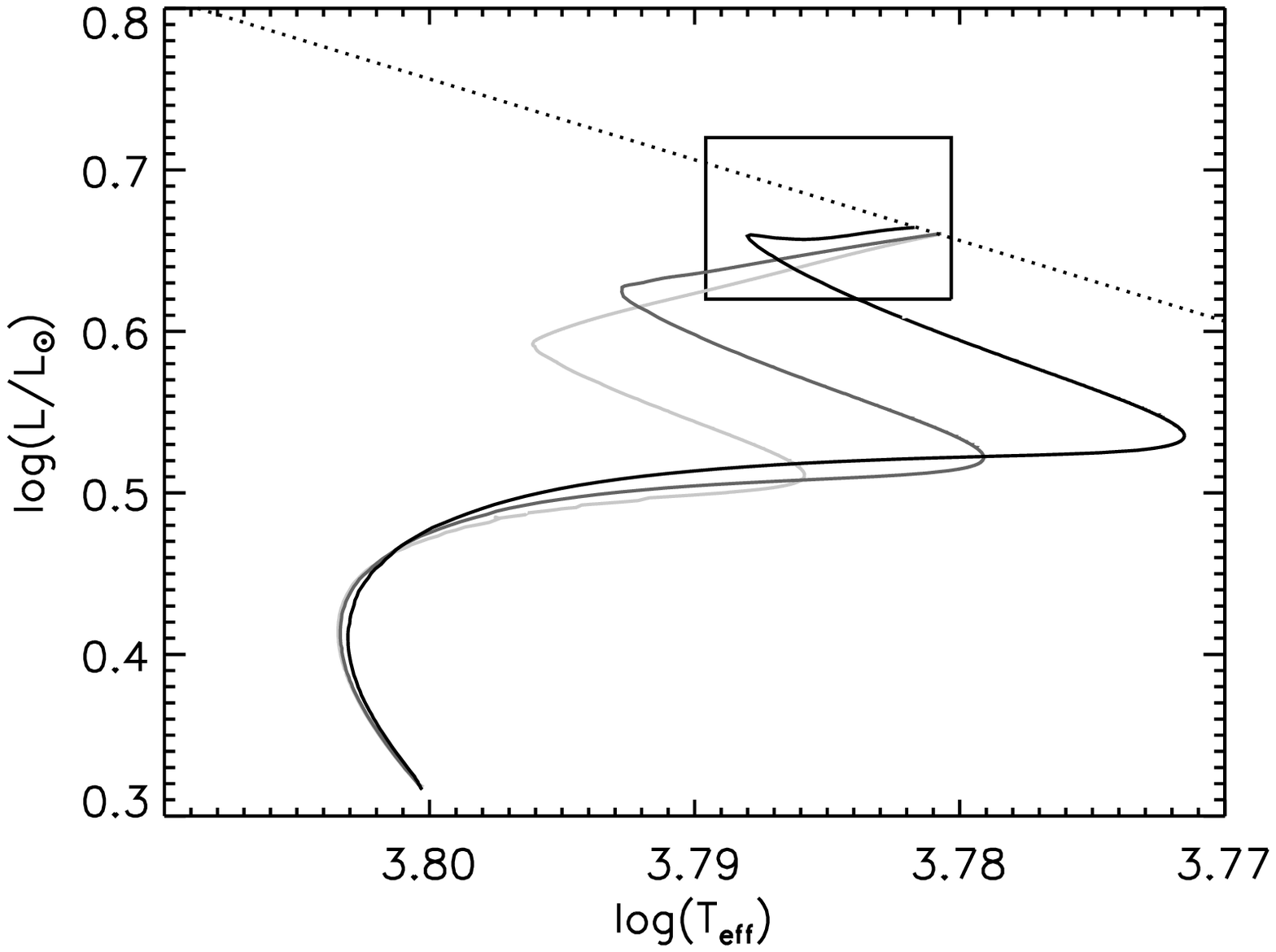}
\includegraphics[width=8cm]{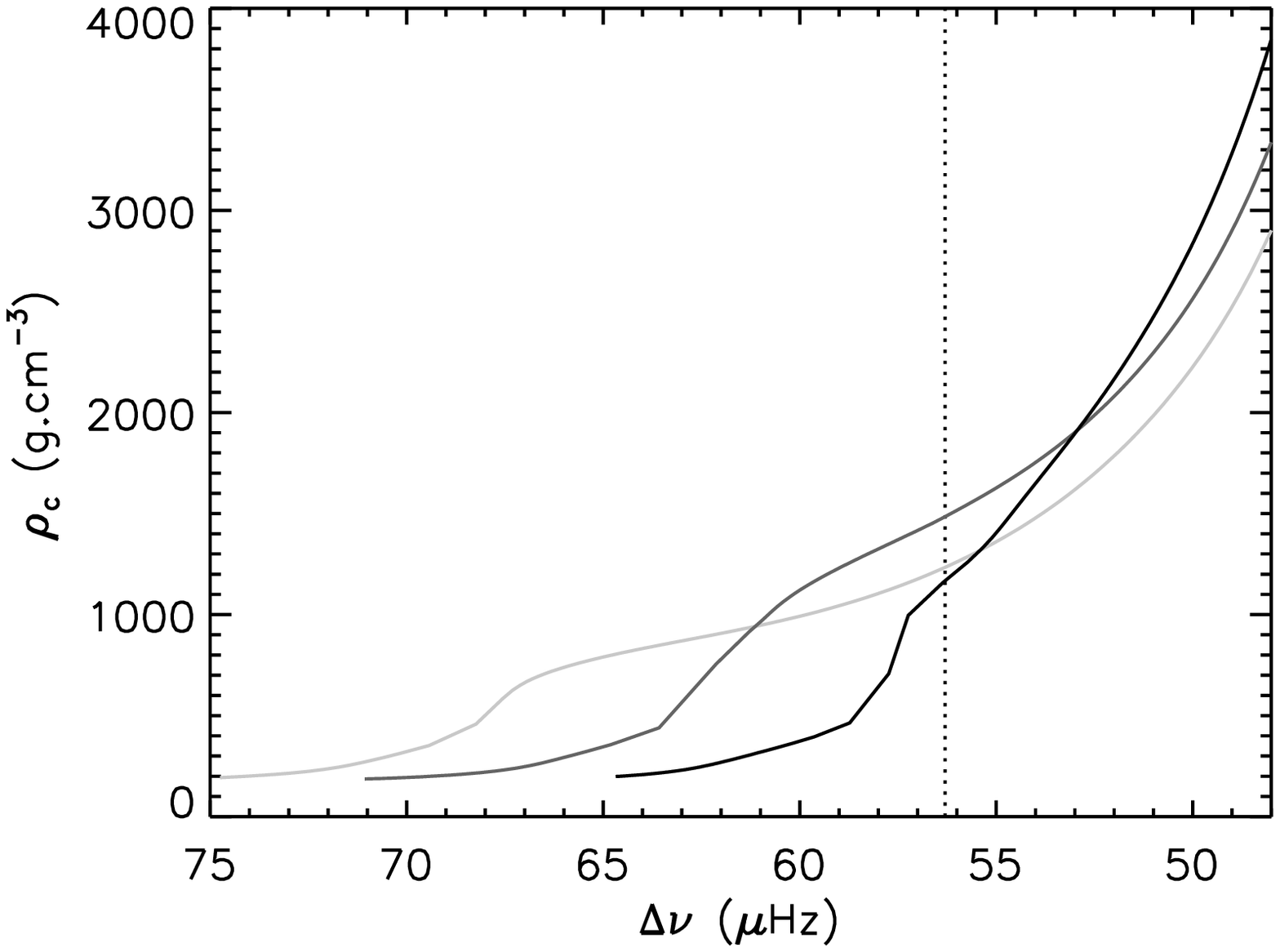}
\end{center}
\caption{Evolutionary tracks in the HR diagram (\textit{left}) and evolution of the central density $\rho\ind{c}$ (\textit{right}) for models with the same mass but different values of $\aov$ ($\aov=0$, 0.08 and 0.15 from light gray to dark gray). The observed mean value of the large separation is represented by the dotted line.
\label{fig_rhoc_ov}}
\end{figure*}

%\begin{figure*}
%\begin{center}
%\includegraphics[width=7.5cm]{fig_hr_y0.ps}
%\includegraphics[width=7.5cm]{fig_nuac_y0.ps}
%\end{center}
%\caption{\textbf{Left}: Effect of a change of the mixing length parameter $\alpha\ind{conv}$ in model $\mathcal{M}_0$ (see text) on the mass $\tilde{M}$ which verifies condition $\mathcal{C}$. \textbf{Right}: Evolutionary tracks in the HR diagram of model $\mathcal{M}_0$ (mass $\tilde{M}_0$ and $\alpha\ind{conv}=0.56$, blue line), a model computed with the same mass but $\alpha\ind{conv}=0.60$ (red dashed line) and 
%%Dashed lines (resp. full lines) represent the evolutionary tracks of models for which $\alpha\ind{conv}=0.56$ (resp. $\alpha\ind{conv}=0.60$). 
%\label{fig_nuac_y0}}
%\end{figure*}

By modifying the value of the initial helium abundance in model $S_0$, we show that $\tilde{M}$ linearly decreases with $Y_0$ (see Fig. \ref{fig_nuac_conv}c).

Like in the previous section, we try to explain this decrease. We search for the new mass $\tilde{M}_2$ that verifies condition $\mathcal{C}$ when we increase the helium abundance to $Y_2>Y_0$. If we keep the same mass $\tilde{M}_0$ (model $S\ind{2a}$), the mean molecular weight $\mu$ increases due to the increase of the helium content. It can be shown from homology relations that if we keep the mass constant and modify $\mu$, the stellar luminosity scales as $L\propto\mu^4$ and the radius scales as $R\propto\mu^{z_2}$ where the exponent $z_2$ depends on the mode of energy generation inside the star (see e.g. \citealt{kippenhahn}). Subgiants are generally burning hydrogen through the CNO cycle and we then have $z_2\simeq0.6$. In this case, combining these two relations with the fact that $L\propto R^2T\ind{eff}^4$, we obtain that $T\ind{eff}\propto L^{5.7}$. Increasing the helium abundance in a model while maintaining the mass constant induces a translation of its evolutionary track in the HR diagram to the left along a line of slope about $5.7$. Since the slope is bigger than that of the iso-$\Delta\nu$ line, the models with higher helium abundance are closer to the TAMS when they reach $\Delta\nu\ind{obs}$ (green dashed line in Fig. \ref{fig_nuac_conv}d). This effect is enhanced by the fact that when increasing $Y_0$, the iso-$\Delta\nu$ line is translated to the left. To reproduce the observed value of $\nuac$ it is therefore necessary to decrease the mass and we have $\tilde{M}_2<\tilde{M}_0$ (model $S\ind{2b}$).

\subsubsection{Dependance of $\tilde{M}$ on the metallicity \label{sect_nuac_zsx}}

%\begin{figure*}
%\begin{center}
%\includegraphics[width=7.5cm]{fig_hr_zsx.ps}
%\includegraphics[width=7.5cm]{fig_nuac_zsx.ps}
%\end{center}
%\caption{\textbf{Left}: Effect of a change of the mixing length parameter $\alpha\ind{conv}$ in model $\mathcal{M}_0$ (see text) on the mass $\tilde{M}$ which verifies condition $\mathcal{C}$. \textbf{Right}: Evolutionary tracks in the HR diagram of model $\mathcal{M}_0$ (mass $\tilde{M}_0$ and $\alpha\ind{conv}=0.56$, blue line), a model computed with the same mass but $\alpha\ind{conv}=0.60$ (red dashed line) and 
%%Dashed lines (resp. full lines) represent the evolutionary tracks of models for which $\alpha\ind{conv}=0.56$ (resp. $\alpha\ind{conv}=0.60$). 
%\label{fig_nuac_zsx}}
%\end{figure*}

Fig. \ref{fig_nuac_conv}e shows that the mass $\tilde{M}$ linearly increases with the abundance of heavy elements $[Z/X]$. 

%We now study the variations of the mass which verifies condition $\mathcal{C}$ when changing the abundance of heavy elements. With a metallicity of $[Z/X]_0$, we have found the mass $\tilde{M}_0$ and the age $\tilde{\tau}_0$ which satisfy $\mathcal{C}$. We now increase the metallicity to $[Z/X]_3>[Z/X]_0$.

To explain this increase, we consider a metallicity $[Z/X]_3>[Z/X]_0$ and we search for the new mass $\tilde{M}_3$. If we keep the mass $\tilde{M}_0$ of model $S_0$ (model $S\ind{3a}$), the main effect of an increase of the metallicity is that the amount of bound-free absorption, which comes from metals, is higher. This leads to an increase of the opacity. In order to study its effect, we have computed homology relations, following exactly the same steps as in Chap.~20 of \cite{kippenhahn}, but considering the opacity $\kappa$ as a basic parameter instead of the mean molecular weight $\mu$. We found that if we keep the mass constant while increasing the opacity, the evolutionary track in the HR diagram is translated to the right on a line of slope about $3.6$ (we note that the models indeed show a translation to the right, but with a somewhat larger slope a little below 5, see purple dashed line in Fig. \ref{fig_nuac_conv}f). Normally, this should bring the models closer to the TAMS when they reach the observed value of the large separation. However, we observe in Fig. \ref{fig_nuac_conv} that the iso-$\Delta\nu$ line is shifted to the right when increasing the opacity and the models with higher metallicity are in fact further from the TAMS (purple full line in Fig. \ref{fig_nuac_conv}f). We therefore need to increase the mass so that the frequency of the avoided crossing is correctly reproduced, i.e. $\tilde{M}_3>\tilde{M}_0$ (model $S\ind{3b}$). 

\subsubsection{Dependance of $\tilde{M}$ on the amount of overshooting \label{sect_nuac_ov}}

\begin{figure}
\includegraphics[width=8.5cm]{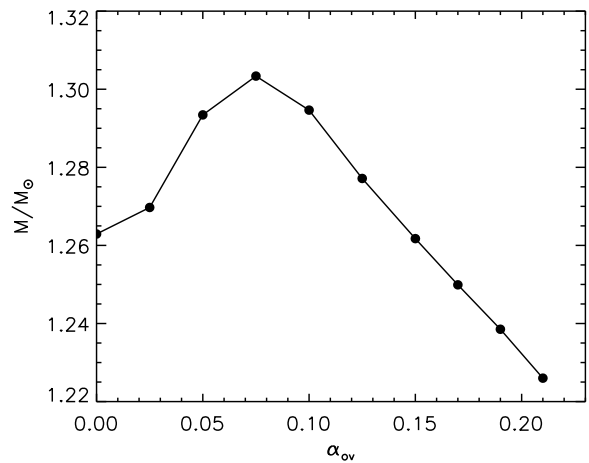}
\caption{Variations of the optimal mass $\tilde{M}$ with the overshooting coefficient $\aov$. Two regimes can be distinguished: for a strong overshooting, the models are close to the TAMS and effect (i) dominates ($\aov>0.1$) whereas for a milder amount of overshooting, effect (ii) takes over ($\aov<0.1$).
\label{fig_mass_ov}}
\end{figure}

We now focus on the influence of overshooting on the mass $\tilde{M}$, which is more complex than the influence of the stellar parameters previously studied. Starting from model $S_0$, we increased the amount of overshooting and determined the mass $\tilde{M}$. Fig. \ref{fig_mass_ov} shows that $\tilde{M}$ increases with $\aov$ for low overshooting coefficients ($\aov<0.06$). Above this limit, the mass $\tilde{M}$ starts to decrease as a function of $\aov$. This decrease eventually becomes linear for $\aov>0.1$. We note that the shape of $\tilde{M}(\aov)$ does not qualitatively change if we consider initial models other than $S_0$ (even though the value of the transitional $\aov$ does vary).

Core overshooting is not supposed to play a direct role since the models we here consider are in the PoMS stage and their convective core has vanished. However, it has an impact on the g-mode frequencies through its influence on the past evolution of the star and the chemical composition in the center. Adding overshooting has two opposing effects on the frequency of the avoided crossing:
\begin{enumerate}
\renewcommand{\labelenumi}{(\roman{enumi})}
\item First, by extending the mixed region associated to the convective core, overshooting increases the size of the hydrogen reservoir available for the nuclear reactions in the center. As a result, the star stays longer in the MS stage, as clearly appears in Fig. \ref{fig_rhoc_ov}. Consequently, for a given fixed mass, an increase of the amount of overshooting brings the star closer to the TAMS when it reaches the observed value of the large separation. Based on what has been said before, we expect the frequency of the avoided crossing to decrease when the amount of overshooting increases. 
%This means that the mass of the model which verifies condition $\mathcal{C}$ should decrease with increasing $\alpha\ind{ov}$. It is indeed the case when the star is close enough to the TAMS, i.e. for large values of the overshooting coefficient. We can see in Fig. \ref{fig_mass_ov} that for $\aov>0.1$, the mass $\tilde{M}$ decreases almost linearly as a function of $\aov$. We note that above a certain amount of overshooting, the models reach the observed value of the large separation while they are still in the MS. In this case, the models are not evolved enough to have avoided crossings in the frequency domain of the observations, as we mentioned before. This can be used to set an upper limit on $\aov$.
\vspace{0.1cm}
%\item When the star is further away from the TAMS, i.e. with milder amounts of overshooting, 
\item But also, adding overshooting increases the size of the mixed core during the MS stage and therefore, when it reaches the TAMS, its helium core is bigger. The layers where hydrogen remains in the star are less deep and the temperature in these regions is thus smaller. The contraction of the star during the PoMS stage needs to be more severe to trigger nuclear burning in these shells. As a result, the central density increases faster with overshooting than it does without overshooting, as can be seen in Fig. \ref{fig_rhoc_ov}. Since the frequency of the avoided crossing is tightly connected to the central density (see Sect. \ref{sect_nuac_age}), we expect $\nuac$ to grow when $\aov$ increases.
%Based on Eq. \ref{eq_BV_rhoc}, we deduce that the \vaisala\ frequency increases more rapidly during the PoMS when overshooting is added. When the star is sufficiently close to the TAMS, this effect has little influence and the first effect is dominant, which explains the behavior of $\tilde{M}$ for large values of $\aov$. However, when the star is evolved enough along the PoMS, i.e. with milder amounts of overshooting, the second effect prevails and $\nuac$ increases with $\aov$. Based on the results obtained in Sect. \ref{sect_nuac_mass}, this implies that the mass $\tilde{M}$ should increase with $\aov$. It is the case for $\aov<0.06$ in Fig. \ref{fig_mass_ov}.
%, which implies that \textit{$\nuac$ should increase with $\aov$}. This second effect is opposite to the first one and becomes predominant when the star is evolved enough along the PoMS. We can see in Fig. \ref{fig_mass_ov} that the mass $\tilde{M}$ increases with $\aov$ for $\aov<0.1$.
\end{enumerate}

In fact, these two effects play a different role, depending on the amount of overshooting. For a mild overshooting, the stars are evolved enough after the TAMS when they reach $\lsmean\ind{obs}$ so that the second effect prevails (see Fig. \ref{fig_rhoc_ov}). When increasing the amount of overshooting of model $S_0$ from $\aov=0$ to $\aov=0.08$, $\nuac$ increases. This means that the stellar mass needs to be increased to ensure that condition $\mathcal{C}$ is satisfied. This accounts for the increase of $\tilde{M}$ at low $\aov$.

On the other hand, for higher amounts of overshooting, the stars are very close to the TAMS when they reach $\lsmean\ind{obs}$. The central density has not had time to increase enough and the first effect is dominant. Fig. \ref{fig_rhoc_ov} indeed shows that if we further increase the overshooting to $\aov=0.15$, the central density decreases and so does the frequency of the avoided crossing. Therefore, the mass needs to be decreased to satisfy condition $\mathcal{C}$.

%In fact, these two effects play a role. For a mild overshooting, the star is far from the TAMS when it reaches $\lsmean\ind{obs}$ and the first effect has little influence. We can see in Fig. \ref{fig_nuac_ov} that $\nuac$ increases with $\aov$ for low values of $\aov$. This tendency is reversed for larger values of $\aov$. Indeed, with a larger amount of overshooting, the star is very close to the TAMS and the first effect prevails, leading $\nuac$ to decrease with $\aov$. 
We note that due to the competition between these two effects, certain masses lead to two different models fulfilling condition $\mathcal{C}$: one with a mild overshooting away from the TAMS, and one with a larger $\aov$, close to the TAMS. For instance for the physics chosen in Fig. \ref{fig_mass_ov}, a mass of $M=1.263\,M_{\odot}$ can satisfy condition $\mathcal{C}$ both without overshooting ($\aov=0$) and with an overshooting of $\aov=0.15$. This observation shows that for subgiants in the early PoMS stage, knowing the frequency of an avoided crossing is not enough to efficiently constrain the amount of core overshooting in a star. 

\section{Application to \cible: model optimization \label{sect_optim}}

\subsection{Computation of two grids of models}

We computed two grids of models, one assuming the mixture of \cite{grevesse93} (furhter noted GN93) and the other the more recent mixture of \cite{asplund05} (further noted AGS05). The grids are built with varying values of the mixing length parameter $\alpha\ind{conv}$, the helium abundance $Y_0$, the metallicity $[Z/X]$ and the amount of overshooting $\alpha\ind{ov}$. The ranges of studied values for each of these parameters as well as the chosen steps are reported in Table \ref{tab_grid}. For each point in the grids, an optimization such as the one described in Sect. \ref{sect_searchC} was performed to determine the stellar mass $\tilde{M}$ and age $\tilde{\tau}$ which satisfy condition $\mathcal{C}$. We then computed the $\chi^2$ function defined by Eq. \ref{eq_chi2}.

\begin{table}
  \centering
  \caption{Ranges and steps adopted for the free parameters of the computed grids of models.
  \label{tab_grid}}
\begin{tabular}{ c  c  c }
\hline \hline
Parameters & Range & Step \\
\hline
\T $Y_0$ & 0.24 to 0.28 & 0.01  \\
$\alpha\ind{conv}$   & 0.48 to 0.72 & 0.04 \\
$[Z/X]$ (dex) & 0.04 to 0.14 & 0.05 \\
\B $\alpha\ind{ov}$ & 0.00 to 0.20 & 0.025 \\
\hline
\end{tabular}
\end{table}

\subsection{Results of the fit \label{sect_results_fit}}

%\begin{table*}
%\centering
%\caption{Values of the stellar parameters obtained from the computation of our two grids of models. 
%%The error bars are deduced from Fig. \ref{fig_chi2_tot} (see text). 
%\label{tab_results}}
%\setlength{\tabcolsep}{10pt} % narrow table: default is tabcolsep = 6pt
%\begin{tabular}{ l | c c | c c }
%\hline \hline
%\T \B & \multicolumn{2}{| c |}{GN93} & \multicolumn{2}{c}{AGS05} \\
%\hline
%\T & low $\aov$ & high $\aov$ & low $\aov$ & high $\aov$ \\
%\B $\aov$ & $0.05^{+0.01}_{-0.05}$ & $0.19 \pm 0.01$ & $0.00^{+0.01}$ & $0.19 \pm 0.01$ \\
%\hline
%\T $M/M_{\odot}$ & $1.285\pm 0.017$ & $1.231\pm 0.012$ & $1.260\pm 0.010$ & $1.224\pm 0.007$ \\
%Age (Gyr) & $4.90 \pm 0.13$ & $5.04 \pm 0.18$ & $4.63 \pm 0.11$ & $4.85 \pm 0.11$ \\
%$R/R_{\odot}$ & $1.959\pm 0.009$ & $1.932\pm 0.011$ & $1.944\pm 0.006$ & $1.928\pm 0.004$ \\
%$T\ind{eff}$ (K) & $5870 \pm 30$ & $5860 \pm 50$ & $6120 \pm 30$ & $6120 \pm 40$ \\
%%$\log g$ & $3.9614 \pm 0.0017$ & $3.9548 \pm 0.0014$ & $3.9594 \pm 0.0011$ & $3.9566 \pm 0.0010$ \\
%$\log g$ & $3.961 \pm 0.002$ & $3.955 \pm 0.001$ & $3.959 \pm 0.001$ & $3.957 \pm 0.001$ \\
%$\alpha\ind{conv}$ & $0.54 \pm 0.01$ & $0.54 \pm 0.02$ & $0.55 \pm 0.01$ & $0.58 \pm 0.02$ \\
%\B $Y_0$ & $0.24^{+0.01}$ & $0.26\pm0.01$ & $0.24^{+0.01}$ & $0.26\pm0.01$ \\
%\hline
%\T \B Minimum $\chi^2\ind{red}$ & 2.37 & 3.08 & 2.16 & 1.88 \\
%\hline
%\end{tabular}
%\end{table*}

\begin{table*}
\centering
\caption{Values of the stellar parameters obtained from the computation of our two grids of models. 
%The error bars are deduced from Fig. \ref{fig_chi2_tot} (see text). 
\label{tab_results}}
\setlength{\tabcolsep}{10pt} % narrow table: default is tabcolsep = 6pt
\begin{tabular}{ l | c c | c c }
\hline \hline
\T \B & \multicolumn{2}{| c |}{GN93} & \multicolumn{2}{c}{AGS05} \\
\hline
\T & low $\aov$ & high $\aov$ & low $\aov$ & high $\aov$ \\
\B $\aov$ & $0.05^{+0.01}_{-0.05}$ & $0.19 \pm 0.01$ & $0.00^{+0.01}$ & $0.19 \pm 0.01$ \\
\hline
\T $M/M_{\odot}$ & $1.285\pm 0.017$ & $1.231\pm 0.012$ & $1.264\pm 0.013$ & $1.210\pm 0.021$ \\
Age (Gyr) & $4.90 \pm 0.13$ & $5.04 \pm 0.18$ & $4.88 \pm 0.11$ & $5.10 \pm 0.18$ \\
$R/R_{\odot}$ & $1.959\pm 0.012$ & $1.932\pm 0.011$ & $1.947\pm 0.007$ & $1.917\pm 0.011$ \\
$T\ind{eff}$ (K) & $5870 \pm 40$ & $5790 \pm 30$ & $5940 \pm 40$ & $6080 \pm 60$ \\
%$\log g$ & $3.9614 \pm 0.0017$ & $3.9548 \pm 0.0014$ & $3.9594 \pm 0.0011$ & $3.9566 \pm 0.0010$ \\
$\log g$ & $3.961 \pm 0.002$ & $3.946 \pm 0.001$ & $3.960 \pm 0.002$ & $3.954 \pm 0.002$ \\
\B $\alpha\ind{conv}$ & $0.52 \pm 0.01$ & $0.52 \pm 0.02$ & $0.52 \pm 0.01$ & $0.56 \pm 0.03$ \\
%\B $Y_0$ & $0.24^{+0.01}$ & $0.26\pm0.01$ & $0.24^{+0.01}$ & $0.26\pm0.01$ \\
\hline
\T \B Minimum $\chi^2\ind{red}$ & 2.48 & 2.25 & 2.11 & 1.81 \\
\hline
\end{tabular}
\end{table*}

\begin{figure}
\includegraphics[width=8.5cm]{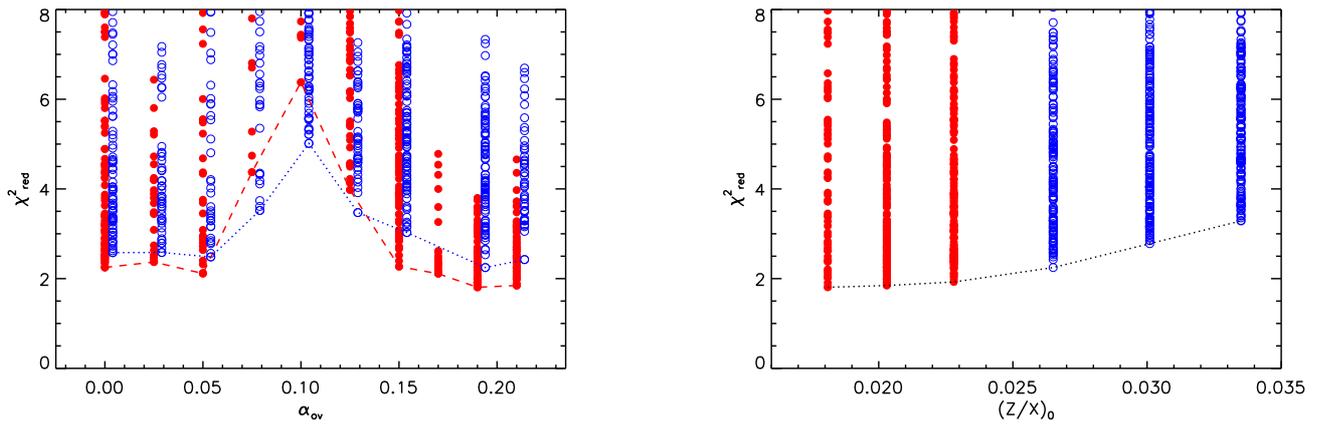} %{fig_ov_bestchi2.ps}
\caption{Values of the reduced $\chi^2$ for the models computed with the mixture of GN93 (blue empty circles) and with the mixture of AGS05 (red filled circles) as a function of the amount of overshooting. For clarity reasons, the symbols of models computed with GN93 were slightly shifted to the right. The best models for each considered value of $\aov$ were linked by a dotted line (GN93) and by a dashed line (AGS05).
%The dots and solid lines give the value of the total $\chi^2$ for models computed with the mixture of GN93 (blue) and the mixture of AGS05 (red). 
%The crosses and dashed lines indicate the value of the $\chi^2$ for the same models omitting the contribution of the frequency of the mode $\ell=1$, $n=12$ (see text). 
%The dashed lines indicate the minimum value of $\chi^2\ind{min}$ of the function and the $\chi^2\ind{min}$ which is used to derive the error bars (see text).
\label{fig_chi2_ov}}
\end{figure}

The first striking comment which can be made about the fit is that the representation of the total $\chi^2$ as a function of the amount of overshooting shown in Fig. \ref{fig_chi2_ov} presents two distinct minima, one for lower values of $\alpha\ind{ov}$ ($\aov<0.05$) and one for moderate values of $\alpha\ind{ov}$ ($\sim 0.19$). This leads us to identify four different families of solutions corresponding to the two different mixtures we considered and either low or high values of $\alpha\ind{ov}$. 

\begin{figure*}
\begin{center}
\includegraphics[width=0.9\textwidth]{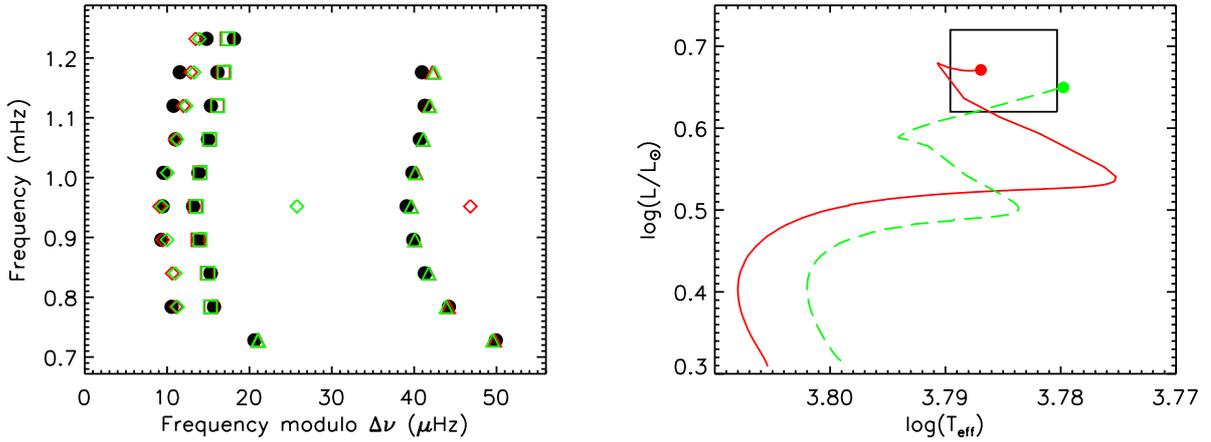}
\end{center}
\caption{\textit{Left:} \'Echelle diagrams of the best models with low overshooting (green) and with high overshooting (red). The squares represent $\ell=0$ modes, triangles $\ell=1$ modes and diamonds $\ell=2$ modes. The observations are represented by the black filled circles. \textit{Right:} Evolutionary tracks in the HR diagram of the two models (green dashed line: low overshooting, red solid line: high overshooting).
\label{fig_echelle_article}}
\end{figure*}

\begin{figure}
\includegraphics[width=8.5cm]{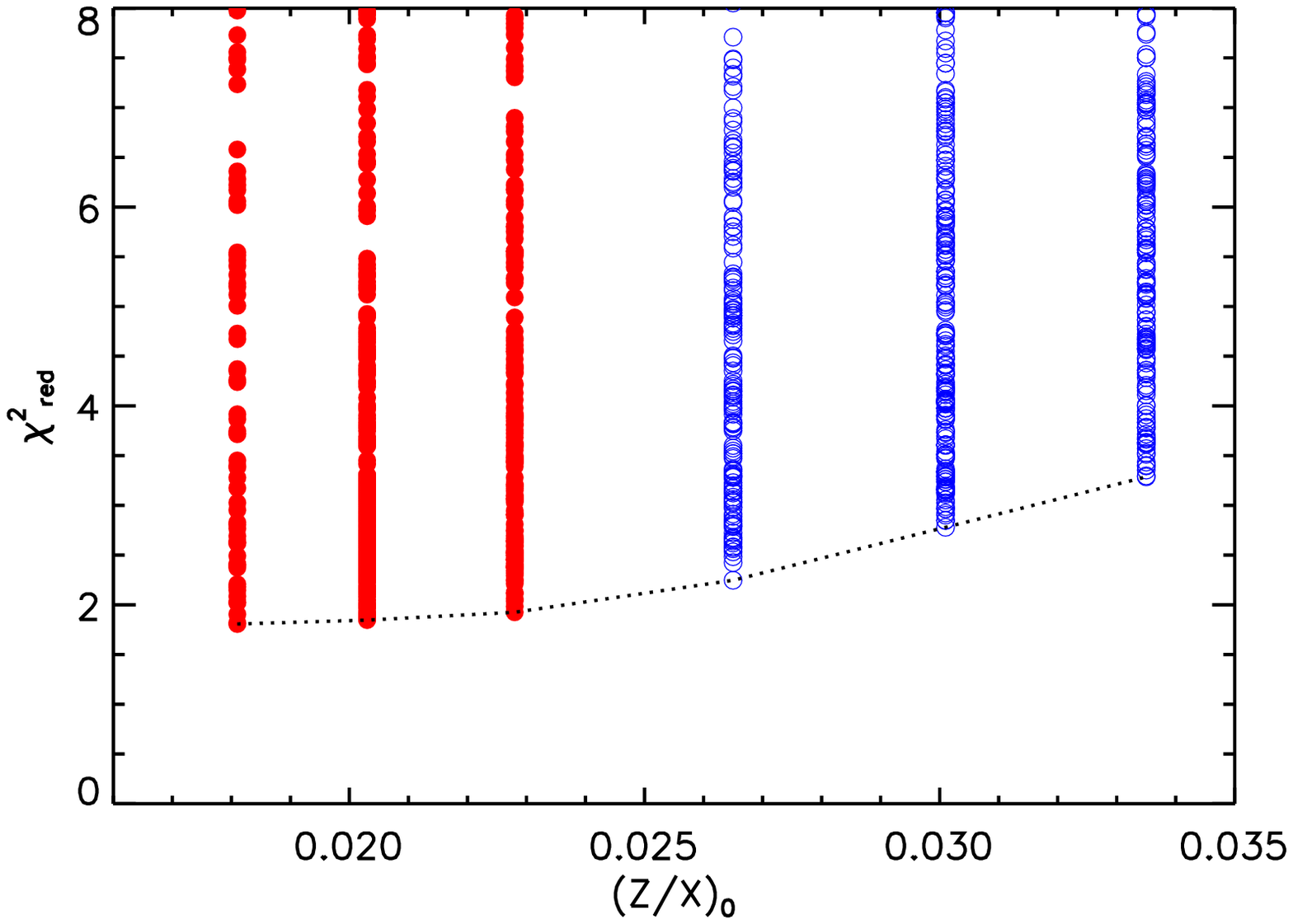} %{fig_ov_bestchi2.ps}
\caption{Values of the reduced $\chi^2$ as a function of the metallicity $[Z/X]$. Symbols are the same as in Fig. \ref{fig_chi2_ov}.
\label{fig_chi2_zsx}}
\end{figure}

We then would like to derive a quantitative interpretation of the $\chi^2$ values we obtain for the models in each of these four families. This is possible since (i) the measured data can be considered to have a Gaussian distribution (\citealt{appourchaux98} showed that for simulated spectra with a resolution equivalent to four months of data, the distribution of the error bars for the central frequencies of the modes is close to a normal distribution) and (ii) for the four families of solutions, the models are roughly linear in their parameters in the regions defined by the uncertainties we will derive for the fitted parameters. This latter condition has been checked a posteriori.
%, for instance Sect. \ref{sect_nuac_conv} to \ref{sect_nuac_zsx} show that the stellar mass varies linearly with $\alpha\ind{conv}$, $Y_0$ and $[Z/X]$ over the whole sample space.

Under these conditions, the goodness of the fit can be estimated by the computation of the reduced $\chi^2$ defined as
\begin{equation}
\chi^2\ind{red} \equiv \frac{\chi^2}{N-P}
\end{equation}
where $N$ is the number of measured data and $P$ the number of free parameters of the fit. More importantly, we can use the values of the $\chi^2$ function to determine a confidence interval for the fitted parameters. Suppose we want to determine the uncertainty in the amount of overshooting $\alpha\ind{ov}$ for instance. For each value of $\alpha\ind{ov}$ in the grid, we determine the set of the other free parameters which minimizes the $\chi^2$ function (the obtained $\chi^2$ are linked by dotted lines in Fig. \ref{fig_chi2_ov}). \cite{numerical} showed that the difference between these $\chi^2$ values and the minimum $\chi^2$ of the whole family of solutions $\chi^2\ind{min}$ defined as
\begin{equation}
\Delta\chi^2 \equiv \chi^2 - \chi^2\ind{min}
\end{equation}
is distributed as a $\chi^2$ with 2 degrees of freedom. This means that the "real" value of the parameter $\alpha\ind{ov}$ has a probability of 68.3\% (resp. 95.4\%, 99.7\%) to belong to the region where $\Delta\chi^2<1$ (resp. $\Delta\chi^2<4$, $\Delta\chi^2<9$). The values of $\alpha\ind{ov}$ for which $\Delta\chi^2=1$ thus give an estimate of the 1-$\sigma$ error in this parameter. This method is applied on all the parameters to determine the uncertainties which are reported in Table \ref{tab_results} for the four different families of solutions.

We thus obtain a precise estimate of the fundamental properties of \cible. The stellar mass is constrained with a precision of about 1\% for each of the four identified families. If we combine them all together, we find a mass of $M=1.25\pm0.05\,M_{\odot}$ for \cible\ (corresponding to a precision of about 4\%). Similarly, we obtain a radius of $R=1.94\pm0.03\,R_{\odot}$ for the star (precision of about 1.5\%). The stellar age is also well constrained: from Table \ref{tab_results}, we derive $\tau=5.02\pm0.26$ Myr (precision of about 5\%). We obtain a $\log g$ of $3.954\pm0.009$ for \cible. We note that the results of the fit are consistent with the first estimates given in Table \ref{tab_fund}. The level of precision we achieve here is to a great extent due to the detection of the $\ell=1$ avoided crossing in the spectrum of \cible\ (see Sect. \ref{sect_discussion}).

As expected, we find that we are also able to obtain information about the interior of \cible. As we mentioned at the beginning of this section, we obtain strong constraints on the amount of overshooting in the star which can be either very low or moderate ($\aov=0.19\pm0.01$). In the latter case, the models are very close to the TAMS. We know that stars are expected to spend very little time in this evolutionary stage, which makes it quite improbable to observe them during this period. Although this point in itself does not justify to ignore these solutions, it means that we should treat them with caution. The best models of the two families we obtained share very similar values of all the observables available for \cible, as shown in Fig. \ref{fig_echelle_article}, so that we cannot discriminate between them. We note however that the models show the existence of an $\ell=2$ mixed mode in the frequency range of the observations, which has different frequencies for the best low-overshooting models (between 973 and 984 $\mu$Hz) and for the high-overshooting ones (between 998 and 1002 $\mu$Hz). If such a mode could be detected in the observed spectrum, we would have a way of discriminating between the two scenarios.  For completeness sake, we checked how the frequencies of the peaks $\pi_2$ and $\pi_3$ of D10 compare with the $\ell=3$ frequencies in our models. In both cases (low and moderate $\aov$), $\pi_2$ and $\pi_3$ appear between 2 and 3 $\mu$Hz higher than the theoretical values (i.e. between 3 and 5 $\sigma$).

An interesting feature of the fit is that the mixing length parameter $\alpha\ind{conv}$ is well constrained for \cible\ and corresponds to a value which is significantly smaller than the solar one, regardless of the family of solution. Combining the families all together, we obtain $\alpha\ind{conv}=0.55\pm0.04$ to be compared to the solar value $\alpha\ind{conv}=0.64$. To strengthen this result, we have checked that we reach similar conclusions when using the traditional mixing length theory instead of the formalism of \cite{canuto96}. In this case, the best agreement is reached for a mixing length of $\alpha\ind{MLT}\sim1.4$, significantly smaller than the solar calibrated value $\alpha\ind{MLT}\sim1.85$.

We also remark that the solutions are in favor of the abundances of heavy elements derived by AGS05. This can be clearly seen in Fig. \ref{fig_chi2_zsx}. This result is strengthened by the fact that even with the mixture of GN93, the best models are found when considering the lowest end of the range of the tested metallicities, i.e. $[Z/X]=0.04$ dex.

These results confirm that the seismic constraints available for \cible\ provide valuable information about the interior of the star. However, they raise a certain number of questions to which the grid-of-model approach in itself cannot answer. Why do we find two distinct families of solutions depending on the value of $\aov$? How come do we manage to constrain the mixing length parameter $\alpha\ind{conv}$ so well in this star? Why do smaller metallicities provide better fits than larger ones? Of what use are the information conveyed by the $\ell=1$ ridge distortion on the coupling between the cavities? We try to address these questions in the following section.

\section{An in-depth analysis of the results of the fit \label{sect_discussion}}

\subsection{Sources of contribution to the $\chi^2$ function \label{sect_contr_chi2}}

For the models computed in Sect. \ref{sect_optim}, we identified three main sources of contribution to the value of the $\chi^2$ function.
%\begin{itemize}
%\item the position of the models in the HR diagram
%\item the oscillation of the mode frequencies due to the bump of $\Gamma_1$ in the second ionization zone of helium
%\item the curvature of the $\ell=1$ ridge due to the coupling between the p-mode and g-mode cavities
%\end{itemize}
%In the next sections, we try to disentangle these contributions to better understand the constraints which they bring about the interior of the star. 
%Our main goal in leading this investigation is in fact to determine to what extent the information we 

%The value of the $\chi^2$ function is dominated by two different contributions: the oscillation due to the bump of $\Gamma_1$ in the second ionization zone of helium and the curvature of the $\ell=1$ ridge due to the coupling between the p-mode and g-mode cavities. In the two next sections, we try to disentangle these contributions to better understand the constraints brought about the interior of the star.

\paragraph{The position in the HR diagram} For each point of the grid, the mass and age are fixed to $\tilde{M}$ and $\tilde{\tau}$, which determines a certain position in the HR diagram. With the abundances of AGS05, the values of $T\ind{eff}$ and $L$ contribute only weakly to the $\chi^2$ of the best models \textbf{(it represents only 0.3\% of the total $\chi^2$ for our best model). The overall fit is thus fully consistent with the spectroscopic constraints.} With the abundances of GN93, the fits point toward an effective temperature about 3-$\sigma$ smaller than the measured one (see Table \ref{tab_results}) which results in a significant contribution to the $\chi^2$ value.

\begin{figure}
\includegraphics[width=8.5cm]{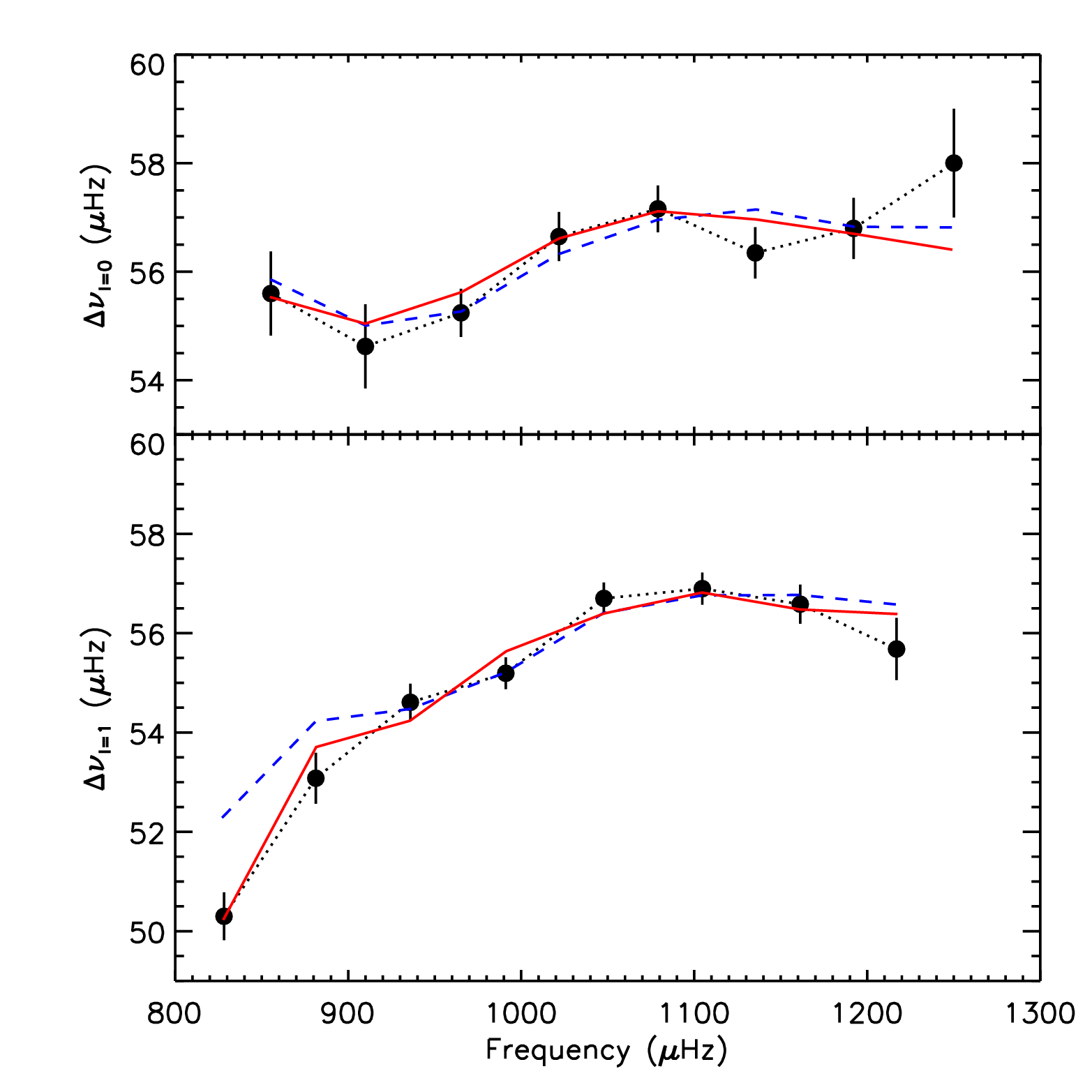}
\caption{\textit{Top}: Profiles of the $\ell=0$ large separations for the observations (black filled circles and dotted lines) and for two models which reproduce well the oscillation due to the $\Gamma_1$ bump. The red solid curve corresponds to the best model obtained in Sect. \ref{sect_optim} and the blue dashed curve to a model computed with $\alpha\ind{conv}=0.68$, $Y_0=0.24$, $\aov=0.10$, $[Z/X]=0.14$ dex, using the GN93 mixture. \textit{Bottom}: Same as the top panel for the $\ell=1$ large separations.
%Comparison between the $\ell=0$ large separation profile of the observations (black filled circles and dashed lines) and two different models which yield a low value of $\chi^2_{\Delta\nu_0}$: model \textbf{A} computed with the GN93 mixture, $\alpha\ind{conv}=0.68$, $Y_0=0.24$, $[Z/X]=0.14$ dex and $\alpha\ind{ov}=0.10$ (blue) and model \textbf{B} computed with the AGS05 mixture, $\alpha\ind{conv}=0.56$, $Y_0=0.26$, $[Z/X]=0.04$ dex and $\alpha\ind{ov}=0.19$ (red).
\label{fig_compare_modeles}}
\end{figure}

\begin{figure*}
\begin{center}
\includegraphics[width=8cm]{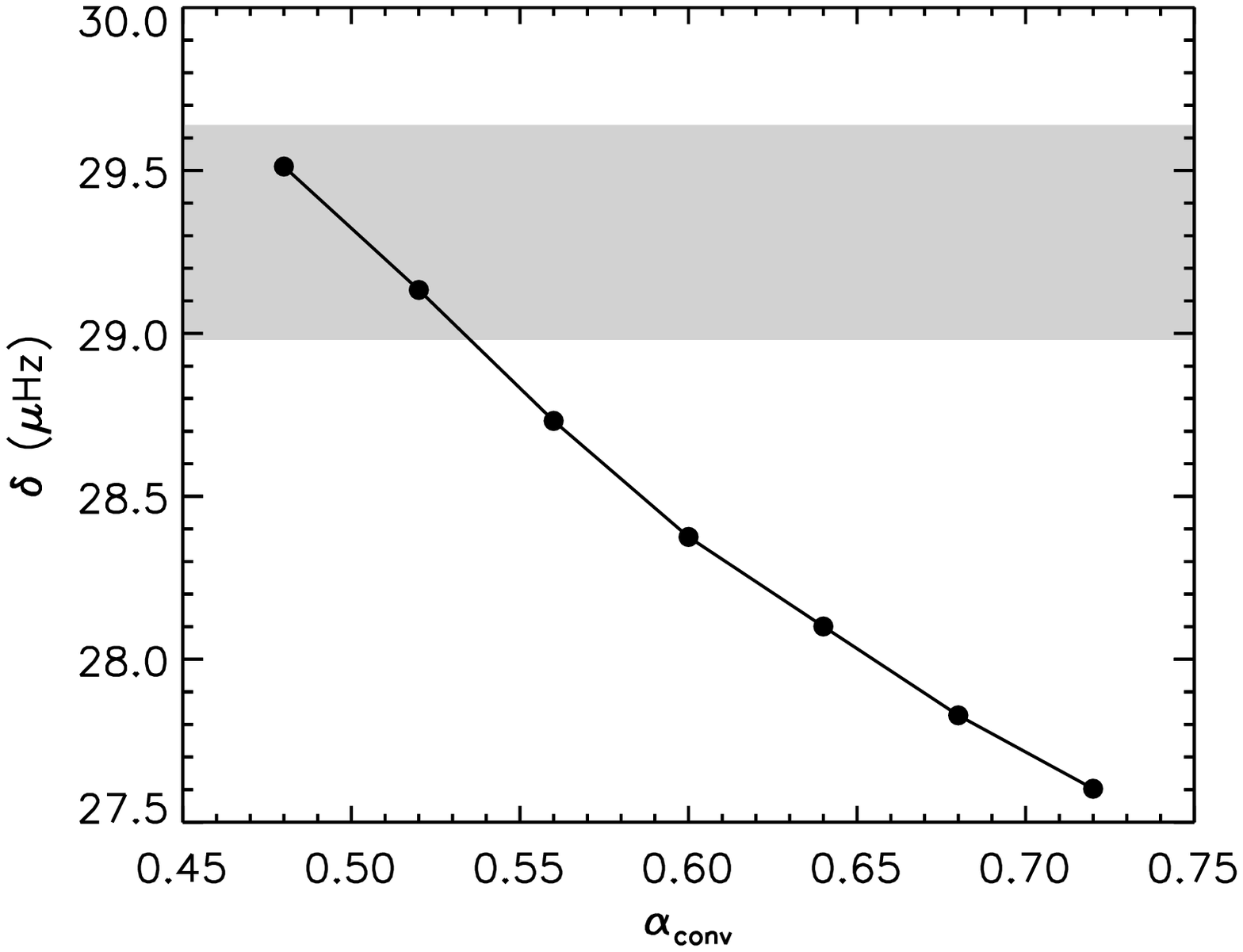}
\includegraphics[width=8cm]{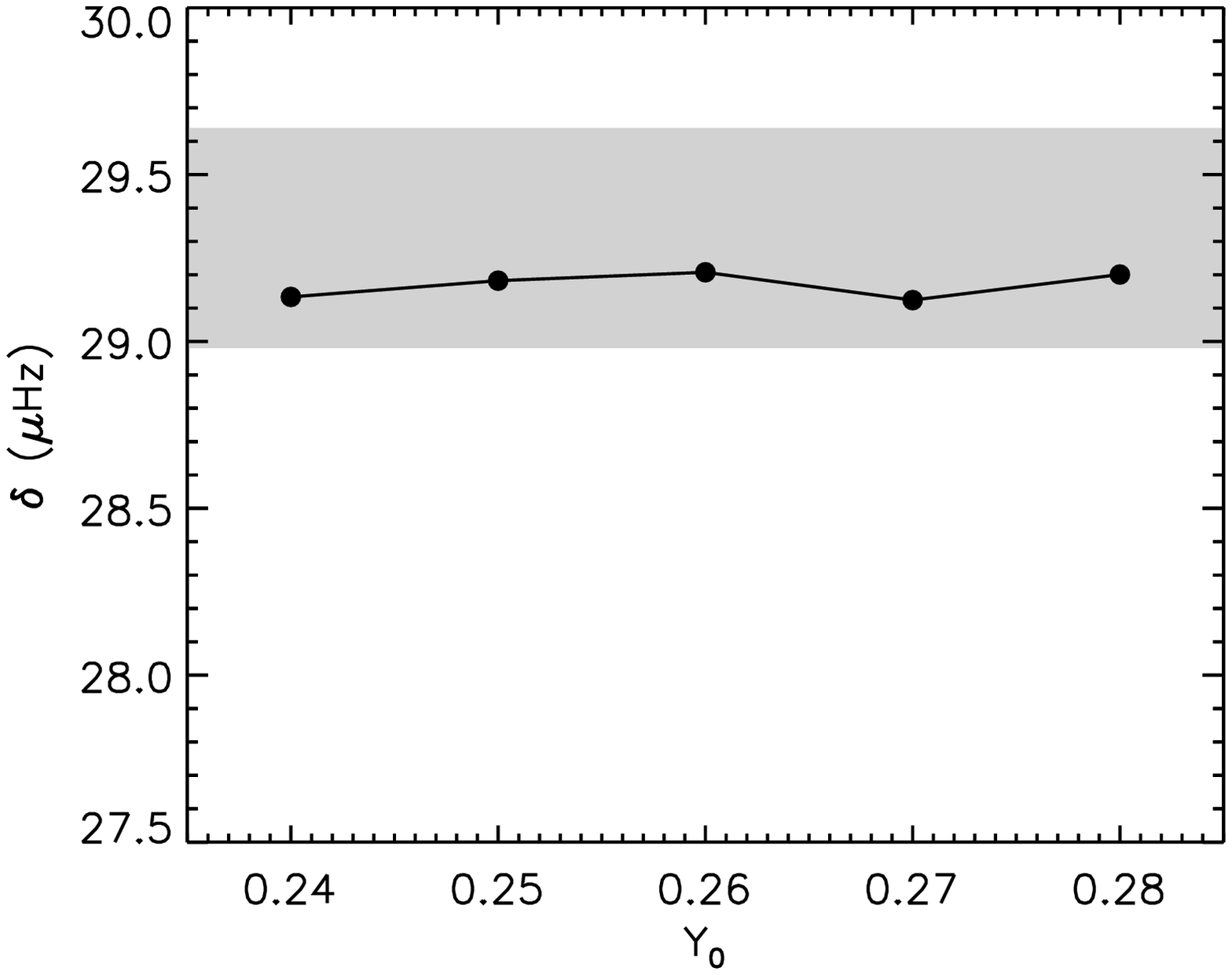}
\includegraphics[width=8cm]{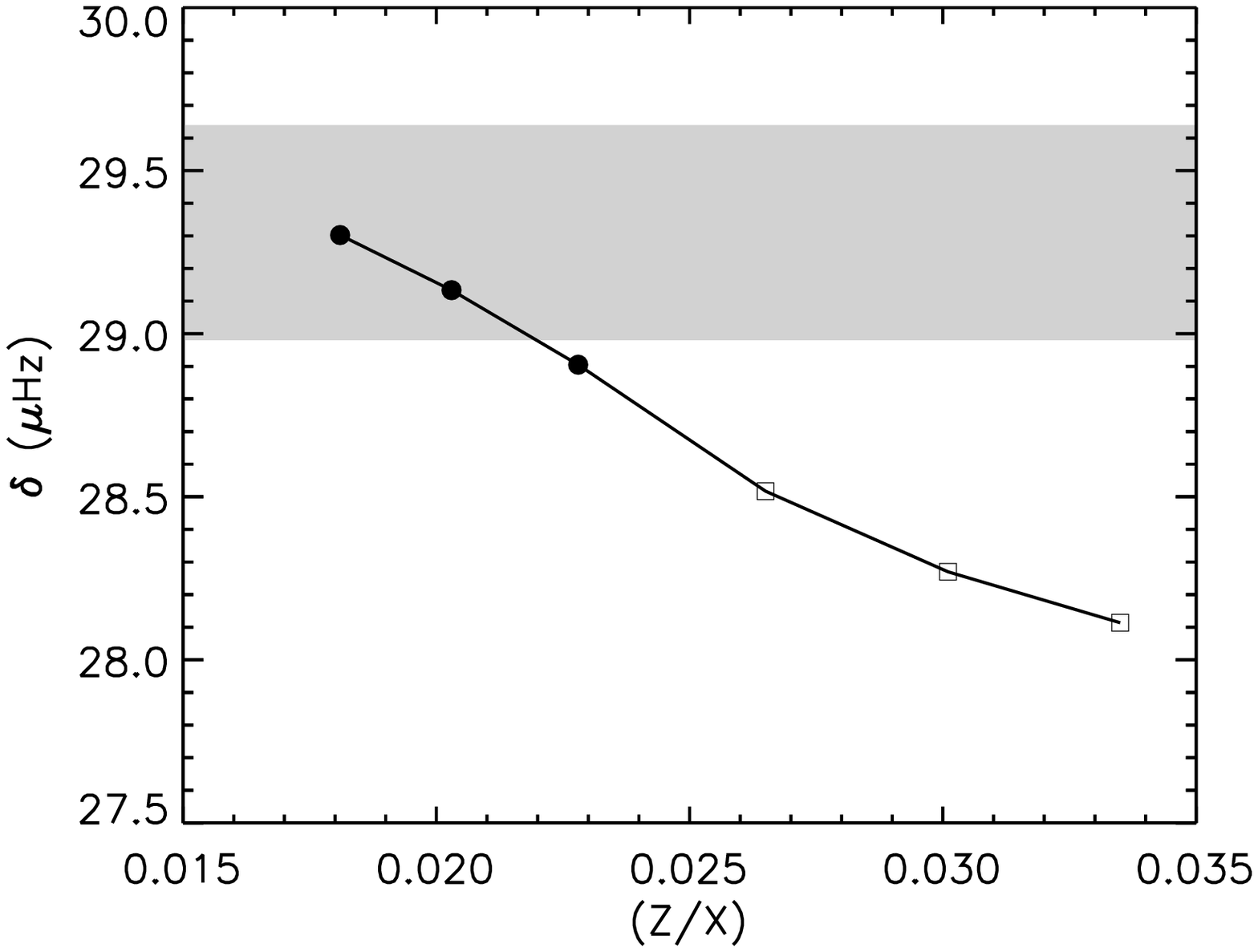}
\includegraphics[width=8cm]{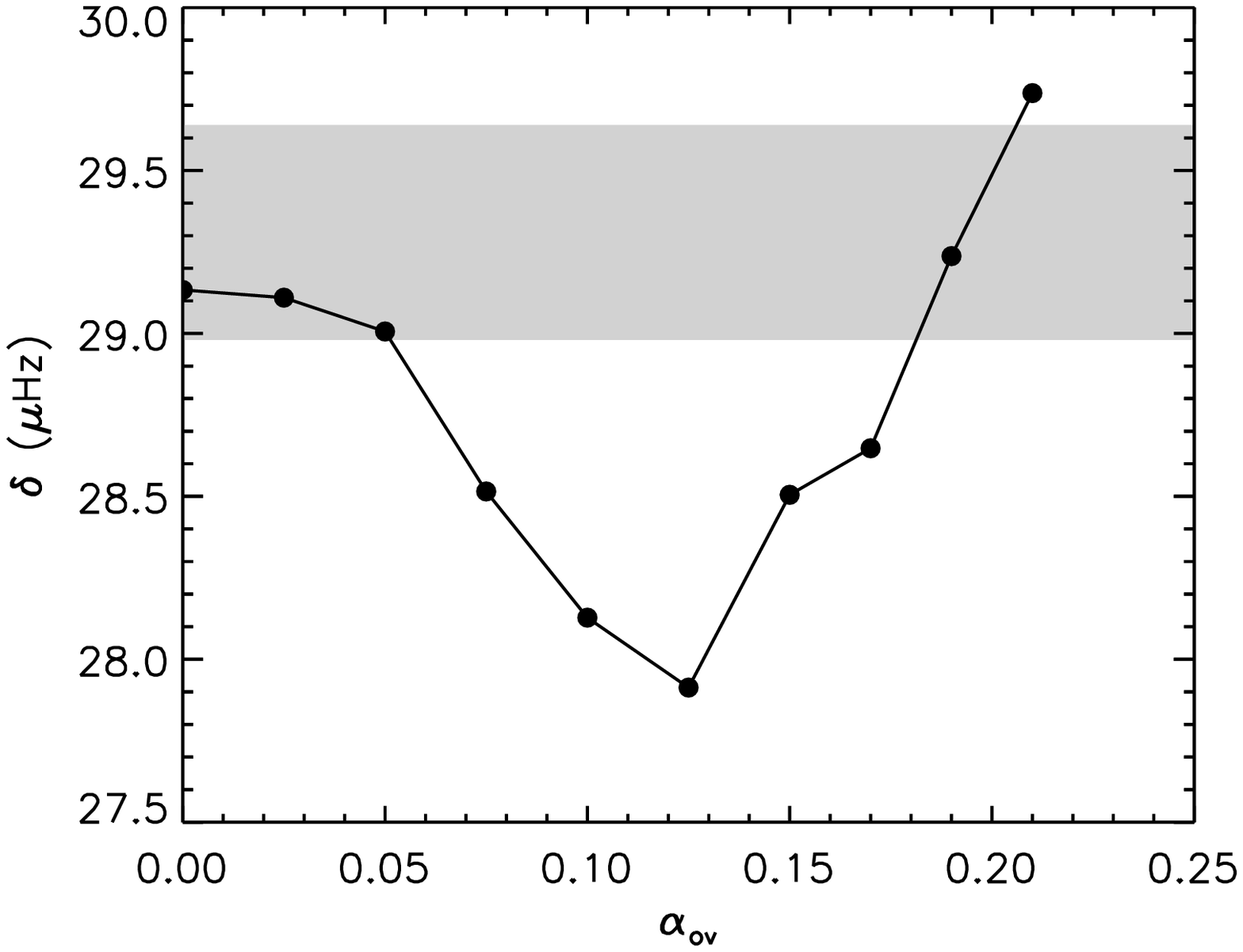}
\end{center}
\caption{Values of the quantity $\delta$ (see text) for models obtained by varying one by one the stellar parameters of the best model without overshooting in Sect. \ref{sect_optim}: the mixing length parameter (\textit{top left}), the initial abundance of helium (\textit{top right}), the metallicity (\textit{bottom left}) and the amount of overshooting (\textit{bottom right}). The filled circles indicate models computed with the abundances of AGS05 and the open squares with those of GN93.
\label{fig_courbure_param}}
\end{figure*}

\paragraph{The oscillation of the frequencies due to the $\Gamma_1$ bump} We have seen in Sect. \ref{sect_sism_constr} that a clear oscillation is observed in the large separation profiles. The models also show an oscillation with a comparable period, which is caused by the bump of $\Gamma_1$ in the helium second ionization zone (HIZ) at an acoustic depth of the order of $\tau\ind{HIZ}\sim0.8$. We note that another oscillation related to the base of the convection zone (BCZ) should also be present with a period around 120 $\mu$Hz ($\tau\ind{BCZ}\sim0.5$), however its amplitude is too low to be detected in the observations. The best agreement with the observed oscillation is obtained for $\tau\ind{HIZ}\sim0.798$, which is consistent with the estimate of the acoustic depth of the glitch we obtained in Sect. \ref{sect_sism_constr} ($\tau\ind{glitch}=0.76\pm0.04$). Fig. \ref{fig_compare_modeles}a shows the comparison between the $\ell=0$ large separation profile of our best model and that of the observations (red curve). We can see that the agreement is good, except at high frequency where the period of the oscillation appears to be smaller in the observations than in the models. However, the observations of p modes over a wider range of radial orders would be required to determine whether or not this decrease of the oscillation period is significant.

Even among the models which satisfactorily fit both the position of the star in the HR diagram and the oscillation of the mode frequencies due to the HIZ, some models yield high values of the $\chi^2$ function. For instance, the model represented in blue in Fig. \ref{fig_compare_modeles} (computed with $\alpha\ind{conv}=0.68$, $Y_0=0.24$, $\alpha\ind{ov}=0.10$, $[Z/X]=0.14$ dex and the mixture of GN93) reproduces the $\ell=0$ large separations quite well but the $\ell=1$ large separation profile shows some striking differences with the observations at low frequency, inducing a high value of the $\chi^2$ function ($\chi^2\ind{red}=7.44$). This shows that the $\ell=1$ ridge for this model is less distorted than it is in the observations. Based on the conclusions drawn in Sect. \ref{sect_analogy}, this means that the coupling between the p-mode cavity and the g-mode cavity is too weak. 

\begin{figure*}
\begin{center}
\includegraphics[width=0.8\textwidth]{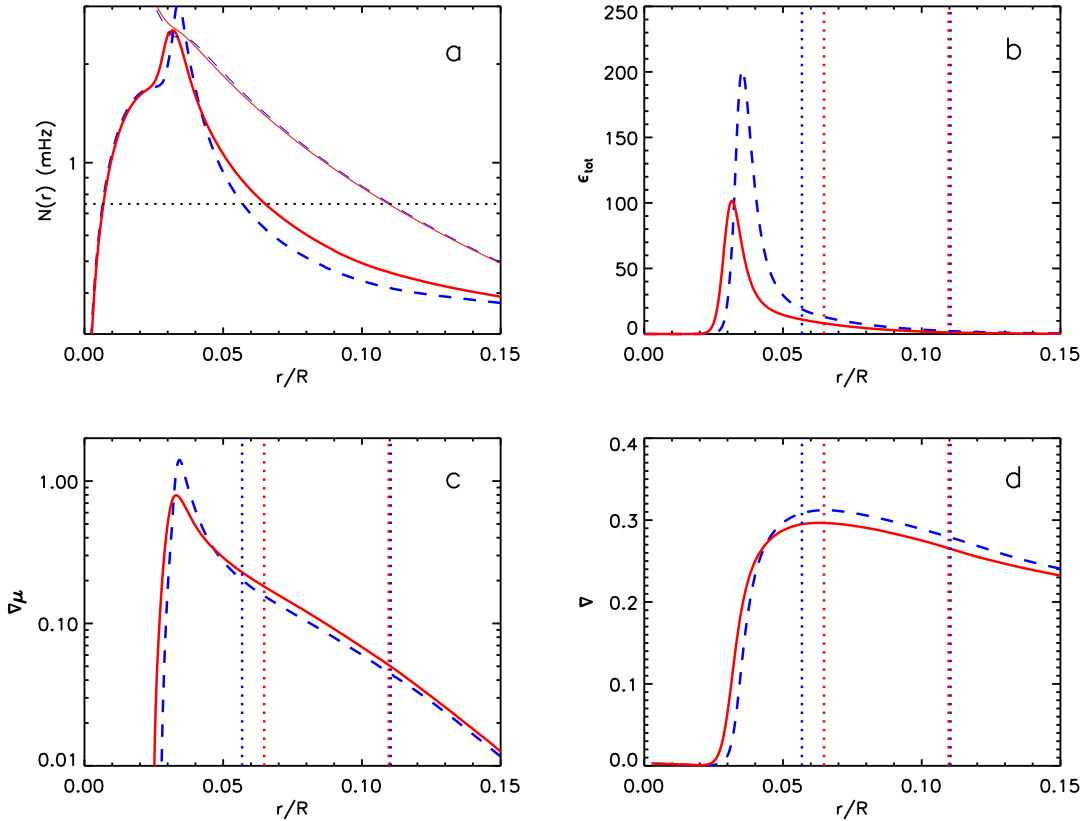}
\end{center}
\caption{Panel \textit{a}: Propagation diagrams of models A (red solid lines) and B (blue dashed lines). The thick lines represent the \vaisala\ frequency, the thin lines the $\ell=1$ Lamb frequency. The dotted line indicates the frequency of the avoided crossing (identical for both models). We then represent the central profiles of the nuclear reaction rate $\varepsilon\ind{tot}(r)$ (panel \textit{b}), of the gradient of mean molecular weight $\gradmu$ (panel \textit{c}) and of the temperature gradient $\nabla$ (panel \textit{d}) for both models. The colored dotted lines indicate the boundaries of the evanescent zone for each model.
\label{fig_courb_mass_ovlow}}
\end{figure*}

%\subsection{Role of the intensity of the $\ell=1$ ridge distortion in constraining the parameters of \cible}
\paragraph{The intensity of the $\ell=1$ ridge distortion} Since the intensity of the $\ell=1$ ridge distortion is a major contribution to the $\chi^2$ function (for instance for the model plotted in blue in Fig. \ref{fig_compare_modeles}, the three lowest frequency $\ell=1$ modes account for 80\% of the $\chi^2$ value) we tried to evaluate its dependence with the free parameters of our fit, in order to understand the results we obtained in Sect. \ref{sect_optim}. For this, we first needed to quantify the intensity of the ridge distortion caused by the avoided crossing in our models. We chose to use the frequency difference $\delta$ of the two modes which most closely surround the avoided crossing. To illustrate why these modes provide a good indicator of the ridge distortion, their frequencies were circled in Fig. \ref{fig_ech_n} for both of the studied cases (weaker coupling and stronger coupling). A low value of $\delta$ indicates a weak coupling between the two cavities and a high value of $\delta$, a stronger one. For \cible, we have
\begin{equation}
\delta = \nu_{1,12} - \nu_{1,11}
\end{equation}
We insist on the fact that the value of $\delta$, and in fact more generally the curvature of the $\ell=1$ ridge, can be used to estimate the strength of the coupling between the cavities \textit{only} if the frequency of the avoided crossing in the models matches the one of the observations (i.e. $\nu_{1,11}=\nu_{\pi_1}$ for \cible). This stresses the great interest of having access to the frequency of the mode which behaves mainly as a g mode in the avoided crossing ($\nu_{1,11}$ in our case). It is also one of the reasons why we focused on finding all possible models satisfying this condition in Sect. \ref{sect_nuac}.

We started from the best model obtained without overshooting in Sect. \ref{sect_optim} and we varied the stellar parameters one by one. The results are shown in Fig. \ref{fig_courbure_param}. Two striking remarks can be made about the obtained plots.

First, we observe that the intensity of the $\ell=1$ ridge distortion can account for most of the results of our fit. The quantity $\delta$ decreases when the mixing length parameter increases, as can be seen in Fig. \ref{fig_courbure_param}a. Only with low values of $\alpha\ind{conv}$ is it possible to reproduce the observed value of $\delta$ ($\delta\ind{obs}=29.31\pm0.33\;\mu$Hz, based on D10). This justifies the fact that our fit points toward a lower mixing length. Similarly, $\delta$ decreases with increasing metallicity. We thus understand why the lower abundances of AGS05 (and more marginally the lowest metallicity with the abundances of GN93) provide a better fit to the observations. The ridge distortion appears to be relatively independent from $Y_0$, and we found in Sect. \ref{sect_optim} that $Y_0$ can hardly be constrained for \cible. Finally, the function $\delta$ depends in a more complex way on the amount of overshooting. Two ranges of $\aov$ satisfactorily reproduce $\delta\ind{obs}$: very low amounts of overshooting ($\aov<0.05$) and moderate amounts of overshooting ($\aov\sim0.2$). This corresponds exactly to what we obtained in Sect. \ref{sect_optim}. All this establishes that the curvature of the $\ell=1$ ridge due to the avoided crossing plays a crucial role in constraining the structure of the interior of \cible.

The second remark which can be made is that the curves of $\delta(\alpha\ind{conv})$, $\delta(Z/X)$ and $\delta(\alpha\ind{ov})$ in Fig. \ref{fig_courbure_param} show a remarkable anti-correlation with the curves of $\tilde{M}(\alpha\ind{conv})$, $\tilde{M}(Z/X)$ and $\tilde{M}(\alpha\ind{ov})$ shown in Fig. \ref{fig_nuac_conv} and \ref{fig_mass_ov}. This is most striking in the case of the variations of $\delta$ and $\tilde{M}$ with the amount of overshooting. This clearly suggests that the coupling between the cavities is inversely proportional to the stellar mass. This would explain why the mass of \cible\ is so well constrained in this study. The next section is dedicated to propose a justification of this phenomenon based on the interior of the computed models.

\subsection{Why are the coupling between the cavities and the stellar mass anti-correlated?}

To understand the relation between the coupling and the stellar mass, we considered two of our models for which the $\ell=1$ ridge distortion is different: model A (computed with $\alpha\ind{conv}=0.48$, $Y_0=0.24$, $[Z/X]=0.09$ dex using the abundances of AGS05) and model B (same as model A but with $\alpha\ind{conv}=0.72$). Model A reproduces well the ridge distortion ($\delta\ind{A}=29.51$, 0.6-$\sigma$ agreement with the observations) contrary to model B ($\delta\ind{B}=27.60$, 5.2-$\sigma$ agreement). Consistent with the results obtained in Sect. \ref{sect_nuac_conv}, model B is more massive than model A ($M\ind{B}=1.37\,M_{\odot}$ and $M\ind{A}=1.20\,M_{\odot}$). We note that the results presented in what follows do not qualitatively change if we consider two other models with conflicting values of $\delta$.

Fig. \ref{fig_courb_mass_ovlow}a shows the propagation diagrams for models A and B. The $\ell=1$ Lamb frequencies of the two models almost overlap in the evanescent zone, but the \vaisala\ frequency of model A in this region is larger than that of model B. This confirms that the coupling between the g-mode cavity and the p-mode cavity is stronger for model A. From Eq. \ref{eq_vaisala}, we know that the \vaisala\ frequency depends on the temperature gradient $\nabla$ and on the gradient of mean molecular weight $\gradmu$. From the end of the main sequence phase to the present age, the profiles of these two quantities have been reshaped by the nuclear reactions in a shell. 

Since model B is more massive than model A, the temperature in its interior is larger, which causes the nuclear reactions of the CNO cycle to be more efficient in the hydrogen burning shell. Fig. \ref{fig_courb_mass_ovlow}b shows that the peak in the nuclear reaction rate $\varepsilon\ind{tot}$ is indeed about twice as strong for model B. As a result, the hydrogen content in the layer where the reactions occur is more severely depleted and the corresponding peak in the gradient of chemical composition is larger. It is therefore logical that in a small region above the hydrogen burning shell (which coincides with the evanescent zone), the gradient $\gradmu$ is smaller for model B. (see Fig. \ref{fig_courb_mass_ovlow}c). Based on Eq. \ref{eq_vaisala}, this suggests that the coupling between the cavities decreases when the stellar mass increases. Similarly, since nuclear reactions are more efficient in higher-mass stars, the ratio $l/m$ (where $l(r)$ is the luminosity going through the shell of radius $r$) is larger for model B and therefore the temperature gradient $\nabla$ is larger in the evanescent zone. This is confirmed by Fig. \ref{fig_courb_mass_ovlow}d. This effect also points toward a weaker coupling for higher-mass stars, according to Eq. \ref{eq_vaisala}.

\subsection{Remark on the effect of microscopic diffusion}

%\begin{figure*}
%\begin{center}
%\includegraphics[width=0.8\textwidth]{fig_gradmu_diff.ps}
%\end{center}
%\caption{\textit{Left}: Propagation diagrams of models with a high overshooting amount ($\aov=0.20$) including microscopic diffusion (purple) and without diffusion (green). \textit{Right}: Profile of the mean molecular gradient $\gradmu$ for these models. The dotted lines indicate the boundaries of the evanescent zone for both models.
%\label{fig_gradmu_diff}}
%\end{figure*}

\begin{figure}
\begin{center}
\includegraphics[width=8.5cm]{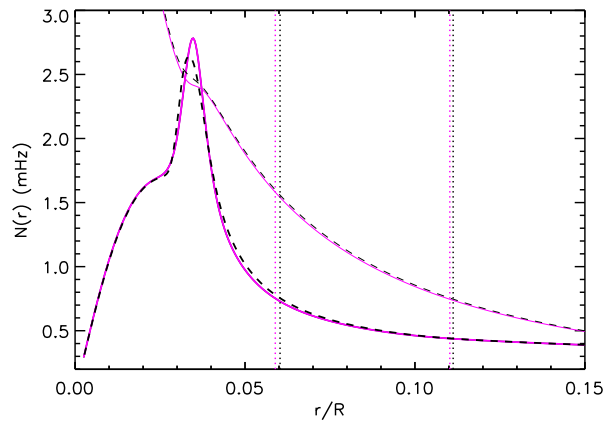}
\end{center}
\caption{Propagation diagrams of two models without microscopic diffusion (black dashed lines) and with diffusion (purple solid lines). The models both satisfy condition $\mathcal{C}$ and share the same values of the free parameters. The symbols are the same as in Fig. \ref{fig_courb_mass_ovlow}.
\label{fig_diag_prop_diff}}
\end{figure}

Microscopic diffusion was not included in the grids of models presented in Sect. \ref{sect_optim}. The main reason is that the time required to compute models including diffusion is several tens of times longer than when it is neglected, which would have severely limited the mesh of our grids. 

However, we have also computed a looser grid of models including microscopic diffusion in order to determine how this process influences our results. For this, we adopted the simplified formulation of \cite{michaud93}. Models were computed with either a low amount of overshooting ($\aov=0$) or a moderate one ($\aov=0.2$), in reference to the two families found in Sect. \ref{sect_optim}. We imposed an initial abundance of heavy elements such that the current surface metallicity matches the observed one. The mixing length parameter was varied from 0.48 to 0.72 (step of 0.04) and the initial helium abundance from 0.24 to 0.28 (step of 0.02). Like before, for every set of parameters, we searched for a model satisfying condition $\mathcal{C}$.

We again obtained a best model offering a very satisfactory match to the observations ($\chi^2\ind{red}=2.1$). All of its characteristics fall into the 1-$\sigma$ error bars obtained without diffusion, except for the age which is smaller (4.1 Gyr for the best model). The main reason why diffusion hardly changes the results is that it appears to have little influence on the coupling between the cavities (which we have shown to play a dominant role in constraining the parameters of the models). This can be seen in Fig. \ref{fig_diag_prop_diff}, which shows the propagation diagrams of two models computed with and without diffusion, both satisfying condition $\mathcal{C}$ and sharing the same values of the free parameters. We observe that the profiles of their \vaisala\ frequency almost overlap. This can be understood as follows. The largest effect of diffusion on the models is that the peak in the profile of $\gradmu$ caused by the withdrawal of the convective core is smoothed. This should have an impact on the coupling since $N\propto\gradmu$. However, this peak is rapidly reshaped by the nuclear reactions in shell. Since the center of the star is radiative, the nuclear reactions are quickly in equilibrium. The abundances of the chemical elements in the core therefore correspond to their equilibrium quantities which do not significantly vary when including diffusion. The nuclear reaction rates thus remain quite similar to the case without diffusion and so does the coupling between the cavities (see Sect. \ref{sect_discussion}).

%Since the center of the star is radiative, the nuclear reactions are in equilibrium. As a result, the abundances of the chemical elements in the center correspond to their equilibrium quantities which do not significantly vary when including diffusion. The nuclear reaction rates thus remain quite similar to the case without diffusion and so does the coupling between the cavities (see Sect. \ref{sect_discussion}). The only major impact of diffusion on the models is that the peak in the profile of $\gradmu$ caused by the withdrawal of the convective core is smoothed. However, this peak is rapidly reshaped by the nuclear reactions in shell which washes out the effects of diffusion.

We note that interestingly, the solutions with high overshooting are found less satisfactory when including microscopic diffusion. This probably has to do with the fact that the models with $\aov=0.2$ which satisfy condition $\mathcal{C}$ are very close to the TAMS. Consequently, the gradient of the mean molecular weight has not yet been reshaped by the nuclear reactions in shell. A deeper analysis, out of the scope of the present paper, would be required to investigate whether or not the inclusion of microscopic diffusion can rule out the solutions with a moderate amount of overshooting.

\section{Conclusion \label{sect_conclusion}}

In this paper, we performed a detailed seismic modeling of the solar-like pulsating subgiant \cible. We first investigated the possibility that the peculiarities observed in the oscillation spectrum of the star by D10 could be due to the existence of mixed modes. By extending a toy-model proposed by \cite{lecturenotesJCD}, we showed that $\ell=1$ avoided crossings involve more than two modes and induce a characteristic and easily recognizable distortion of the $\ell=1$ ridge in the \'echelle diagram. We found PoMS (post main sequence) models with an $\ell=1$ avoided crossing which can account for both the observed curvature of the ridge and the presence of one of the peaks detected outside the identified ridges. On the other hand MS (main sequence) models fail to reproduce these features. We thus established a firm detection of mixed modes in the spectrum of \cible\ as well as the PoMS status of the star. We also pointed out that the curvature of the $\ell=1$ ridge depends on the coupling between the p-mode and g-mode cavities and should therefore bring information about the structure of the star in the evanescent zone which separates the cavities.

We then remarked that the methods usually applied to model stars needed to be adapted to the special case of stars with avoided crossings. We therefore proposed a new approach to the grid-search modeling. We first showed that, for a given physics, the combined knowledge of the frequency of an avoided crossing and of the mean large separation is enough to obtain very precise estimates of the stellar mass and age (which we noted $\tilde{M}$ and $\tilde{\tau}$ in this paper). We then described a method to determine $\tilde{M}$ and $\tilde{\tau}$ in a systematic way in a grid of models. 

This method was applied to gain insights on how the stellar mass $\tilde{M}$ varies with the different stellar parameters and efforts were made to physically understand these variations. To model \cible, we used the proposed method to compute two grids of models, one assuming the heavy element abundances of GN93 (\citealt{grevesse93}) and the other assuming those of AGS05 (\citealt{asplund05}). 
%We remarked that for most of the computed models, the $\ell=1$ ridge is significantly less distorted than it is in the observations, which suggests that the coupling between the cavities is too weak for these models.
In this framework, we were able to strongly constrain the mass of \cible\ ($M=1.25\pm0.05\,M_{\odot}$) and its age ($\tau=5.02\pm0.26$ Myr). The stellar radius was found with a precision as good as 1\% ($R=1.94\pm0.03\,R_{\odot}$). 

We also obtained constraints on the physics of the interior of the star. Two different families of solutions were found, depending on the amount of overshooting that existed above the convective core during the main sequence phase: one with a very low amount of overshooting ($\aov<0.05$), and the other with a moderate amount of overshooting ($\aov=0.19\pm0.01$). The models of the latter family provide the best agreement with the observations, but they are very close to the TAMS and thus in a stage we should be unlikely to observe. The mixing length parameter was found to be significantly smaller than the solar one ($\alpha\ind{conv}=0.55\pm0.04$ compared to $\alpha_{\odot}=0.64$ using the formalism of \citealt{canuto96}). Finally, we showed that the models computed with the revised heavy-element abundances of AGS05 (\citealt{asplund05}) provide a better match with the observations than the abundances of GN93 (\citealt{grevesse93}).

A more advanced study of the models of our grids enabled us to explain the results we obtained about the overshooting, the mixing length and the metallicity for \cible, in terms of stellar structure. We established that the intensity of the $\ell=1$ ridge distortion associated with the observed avoided crossing plays a crucial role in constraining the parameters of the models computed for \cible. We showed that it can account for most of the obtained results. We also observed that the intensity of the coupling between the p-mode and g-mode cavities is strongly anti-correlated to the stellar mass in our models. We therefore suggested that these two quantities might be related to each other. By comparing the models of our grids, we managed to establish the link between them, mainly due to the fact that the temperature and chemical composition profiles (upon which the \vaisala\ frequency depends) strongly depend on the nuclear reaction rate in the hydrogen burning shell, and thus on the stellar mass. We therefore a posteriori understand why the mass is so well constrained in \cible.

The subgiant \cible\ is the first target for which a detailed modeling was led trying to reproduce all the properties of an avoided crossing. It confirmed to a large degree that the detection of mixed modes provides unprecedented opportunities of probing the deep interior of stars. It also gave the opportunity to show that $\ell=1$ avoided crossing have a second interest: they enable us to probe the evanescent zone by yielding an estimate of the intensity of the coupling between the p-mode and g-mode cavities. We found here that this coupling strongly depends on the stellar mass and that we can thus obtain indirect constraints on the internal structure of the star. This shows that subgiants have the potential to contribute to the on-going debates which currently exist about the amount of overshooting at the boundary of convective zones, or the abundances of heavy elements in stars. This is all the more exciting since the space mission Kepler already claimed the detection of $\ell=1$ avoided crossings in several targets (\citealt{chaplin10}, \citealt{metcalfe10}).

\begin{acknowledgements}
This work was supported by the Centre National d'Etudes Spatiales (CNES).
\end{acknowledgements}

\begin{appendix}

\section{Location of iso-$\Delta\nu$ regions in the HR diagram \label{app_dn}}

We investigate the location in the HR diagram of models which have the same large separation $\Delta\nu$ but different masses. Let us consider two such models with masses $M$ and $M'$. Since the large separation is proportional to the square root of the mean density in the star, we have
\begin{equation}
\left(\frac{\Delta\nu}{\Delta\nu'}\right) \sim \left(\frac{M}{M'}\right)^{1/2} \left(\frac{R}{R'}\right)^{-3/2} \sim 1 \label{eq_app1}
\end{equation}
It has been observationally established that there exists a very tight relation between the large separation and the frequency of maximum power of the oscillations $\nu\ind{max}$. \cite{stello09} have found $\Delta\nu\propto\nu\ind{max}^{0.77}$. We gather from this result that models which have the same large separation also share equivalent values of $\nu\ind{max}$. Following \cite{brown91}, we expect $\nu\ind{max}$ to be proportional to the acoustic cut-off frequency and we have
\begin{equation}
\left(\frac{\nu\ind{max}}{\nu\ind{max}'}\right) \sim \left(\frac{M}{M'}\right) \left(\frac{R}{R'}\right)^{-2} \left(\frac{T\ind{eff}}{T\ind{eff}'}\right)^{-1/2} \sim 1 \label{eq_app2}
\end{equation}
By combining Eq. \ref{eq_app1} and \ref{eq_app2}, we obtain $R\sim T\ind{eff}^{1/2}$. Since the luminosity is such that $L\sim R^2T\ind{eff}^4$, we conclude that the models which have the same large separation satisfy
\begin{equation}
L\propto T\ind{eff}^5
\end{equation}
They are therefore located in the HR diagram on a line of slope equal to 5. This property is very well satisfied by the stellar models computed in this work.

\end{appendix}

% for the bibliography, at the end
\bibliographystyle{aa.bst} % style aa.bst
\bibliography{biblio} % your references Yourfile.bib

\end{document}